# A macro-microscopic coupled constitutive model for fluid-saturated porous media with compressible constituents

Jia-Yu Liang[1, 2, 3], Yue-Ming Li[1, 2*], Erich Bauer[3]

1. State Key Laboratory for Strength and Vibration of Mechanical Structures, Xi'an Jiaotong University, Xi'an, 710049, China
2. Shaanxi Key Laboratory of Environment and Control for Flight Vehicle, Xi'an Jiaotong University, Xi'an, 710049, China
3. Institute of Applied Mechanics, Graz University of Technology, Technikerstrasse 4/II, 8010 Graz, Austria

**Abstract**

The paper provides a macro-microscopic coupled constitutive model for fluid-saturated porous media with respect to the compressibility of the solid skeleton, the real solid material and the fluid phase. The derivation of the model is carried out based on the porous media theory and is consistent with the second law of thermodynamics. In the present paper, two different sets of independent variables are introduced to implement the coupled behavior between the compressibility of the solid skeleton and the real solid material. Altogether the proposed model exploits five independent variables, i.e. the deviatoric part of the right Cauchy-Green deformation tensor, the partial density of solid phase, the density of the real solid material, the density of the real fluid material and the relative velocity of the fluid phase. Subsequently, the linearized version of the proposed constitutive model is also presented and compared with some models by other authors. It is found that Biot's model can also be derived based on the linearized version of the proposed model, which indicates that the present work bridges the gap between the porous media theory and Biot's model. Compared with Biot's model, the present model can provide the evolution of the porosity by considering the volumetric strain of the solid skeleton and the volumetric strain of the real solid material.

**Keywords**: Saturated porous media; Porous media theory; Constitutive relations; Biot's model.

*Corresponding author
Email addresses: ljy@stu.xjtu.edu.cn (Jia-Yu Liang), liyueming@xjtu.edu.cn (Yue-Ming Li), erich.bauer@tugraz.at (Erich Bauer)

# 1. Introduction

The constitutive modeling of fluid-saturated porous media has a wide range of applications, such as in the fields of geotechnical engineering and biomechanics. A saturated porous media is composed of a solid skeleton and a pore space which is filled by a fluid. In this study, both the solid and the fluid constituents are assumed to be compressible. In this context, the difficulty to describe the different phases in a unified manner must be dealt with. Solving this problem requires the analysis of the coupled relationship between the solid deformation and the pore-fluid flow behavior. Usually, there are two main approaches that may be adopted for modeling the coupled behavior between the solid and fluid phases of porous media (Chen and Hicks, 2011; Coussy et al., 1998; Gajo, 2010; Laloui et al., 2003; Lewis and Schrefler, 1998; Rajagopal and Tao, 2005; Serpieri and Travascio, 2017; Wei and Muraleetharan, 2002; Zhang, 2018).

One such approach is based on the pure macroscopic theory, which originated from the classical theory of Terzaghi (1936) for porous media with incompressible constituents and the theory by Biot (1941) for porous media with compressible constituents. In the past few decades, the theory has been extended to finite deformations, inelastic and anisotropic solid materials (Anand, 2017; Biot, 1956; 1977; Gajo, 2010; Hu, 2016; 2017; 2018). This approach has been widely evaluated in different publications (e.g. Carroll and Katsube, 1983; Murrel, 1965; Nur and Byerlee, 1971). The constitutive relations derived by pure macroscopic theory is based on the assumption that a macroscopic thermodynamic potential exists (Biot, 1972; Coussy, 2004; Gajo, 2010).

A different refined approach is based on the theory of porous media, which originated from the work related to the volume fraction introduced by Fillunger (1936). Further developed theories of porous media combines the classical field theory and the concept of volume fraction (Bowen, 1980; 1982). The porous media theory differs from the traditional mixture theory by introducing the volume fraction as a measure of the local portions of the individual materials. The theory provides a rigorous framework where the constituents are described by superposed and interacting continua. Extensive studies have been conducted to investigate the properties of various theoretical models (e.g. Borja, 2005, 2006; Borja and Koliji, 2009; Bluhm and de Boer, 1997; de Boer, 1996, 2000, 2005; de Boer and Bluhm, 1999; de Boer and Ehlers, 1990; Ehlers, 2002, 2009, 2018). For a historical survey of past contributions in this field, the interesting reader is referred to de Boer (1996, 2000) and Ehlers (2018).

Apart from the two approaches mentioned above, there are also some other approaches for the modeling of porous media such as the homogenization theory and the approaches based on micromechanics consideration, e.g. Berryman (2005); Hornung (1997); Lopatnikov and Cheng, 2002; Schanz, 2009).

It is worth noting, however, that Biot's theory is not consistent with the modern theory of porous media (de Boer, 2005; Ehlers, 2018). The discrepancy between the two approaches has been discussed and studied by many researchers (de Boer, 2000, 2005; Lade and de Boer, 1997, Lopatnikov and Cheng, 2004; Schanz, 2009; Schanz and Diebels, 2003). The main difference between the porous media theory and the pure macroscopic theory lies in effective stress law. Lopatnikov and Cheng (2004) pointed out that "The effective stress law interpreted from the theory of porous media, however, found that the micromechanical relation for the effective stress coefficient differed from that of Biot by some porosity factor. Since Biot's relation has been widely supported by laboratory measurements and is used in practice, this gap needs to be bridged, which so far has not been accomplished". Cheng (2016) also stated "Based on an extensive historical study, de Boer called attention to a controversy between Karl von

Terzaghi, father of soil mechanics, and Paul Fillunger, father of mixture theory, on the correct form of the effective stress. According to de Boer, the controversy is still not settled". Although Coussy et al. (1998) has proved that the macroscaled equations derived from porous media theory can be reformulated in terms of the measurable quantities involved in the macroscale theories, it is worth remarking that only some basic concepts in mixture theory are adopted in Coussy et al.'s derivations. The constitutive relations proposed by Coussy et al. is based on Lagrangian porosity rather than Eulerian porosity, which is widely adopted in porous media theory. The gap between the porous media theory and pure macroscopic theory has not been bridged (de Boer, 2005; Ehlers, 2018). For a detailed description of this topic, readers are referred to Lade and de Boer (1997), de Boer (2000, 2005) and Schanz (2009).

The concept of volume fraction plays a crucial role in describing the mechanical behavior of porous media (Lopatnikov and Gillespie, 2009). It has been pointed out that the introduction of the porosity concept enables porous media theory to be able to address the microscopic scale (de Boer, 2005). The influence of constituents volume fractions, however, is not taken into account in Biot's model. Several efforts have been made to address this problem.

In porous media theory, these efforts began with the introduction of an additional balance equation to the formulation of an evolution equation for the volume fraction (Bowen, 1982).Subsequent to this and apart from the mass balance and moment balance equations for the fluid phase, a balance equation of porosity was added in the control equations by Wilmanski (1998). A fundamental idea in the proper treatment of the compressibility of the solid skeleton and of the real solid material is the introduction of a multiplicative split of deformation gradient (Bluhm and de Boer, 1997). Due to the relation among the volumetric strain of the solid skeleton, the volumetric strain of the real solid material and the volumetric strain related to the volume fraction of the solid phase, any two of the three variables can be adopted to illustrate the evolution of porosity (Bluhm and de Boer, 1997). Subsequent developments have addressed that the multiplicative split plays a crucial role in the constitutive modeling of porous media with respect to the evolution of the porosity (de Boer, 2005; Ehlers, 2018).

In pure macroscale theory, Coussy et al. (1998) takes the evolution of porosity into account by using the concept of Langrangian porosity. The volumetric constitutive relation of real solid material is taken into consideration by Gajo (2010) based on the method of the multiplicative split of the deformation gradient. Gajo's model was extended by Hu (2016; 2018) in order to consider the coupled behavior between the volumetric strains of the solid skeleton and the real solid material.

The free energy densities related to the constituents in the porous media theory and pure macroscale theory are all defined at the macroscopic scale. Differing from these results, Lopatnikov and Cheng (2002) established the macroscopic and microscopic constitutive relations by using variational theory at the macroscopic and microscopic scales, respectively. Further, the coupled behavior between the macroscopic and microscopic constitutive relations are taken into account by Serpieri and Rosati (2011) based on the variational macroscopic theory of porous media. In the model by Serpieri-Rosati, a relationship between the macroscopic and microscopic constitutive relations is established. The linearized version of the model for the solid phase can be represented by a symmetric $7 \times 7$ matrix (Serpieri, 2011). There are four volumetric moduli in the Serpieri-Rosati model, however, while three volumetric moduli appear in Biot's model.

The main purpose of the present paper is two-fold. First, we provide a macro-microscopic coupled constitutive model based on the theory of porous media theory. Second, we show that Biot's model can be derived from the linearized version of the proposed model. Therefore the model proposed in the present paper bridges the gap between the porous media theory and Biot's model.

In order to implement the coupled behavior between the compressibility of the solid skeleton and the real solid material, two different sets of independent variables are considered for the Helmholtz free energy function. Altogether five independent variables are used for the proposed model, i.e. the deviatoric part of the right Cauchy-Green deformation tensor, the partial density of solid phase, the density of the real solid material, the density of the real fluid material and the relative velocity of the fluid phase. For modelling the volumetric part of the constitutive relation of the solid material, the constitutive relation of the real solid material is first introduced at the microscopic scale and then coupled with the constitutive relation of the solid skeleton at the macroscopic scale. The linearized version of the proposed constitutive relations, which is convenient for practical application, is also derived. Although the number of material parameters is the same as in Biot's model, the present model also delivers the evolution of the porosity based on the volumetric strain of the solid skeleton and the volumetric strain of the real solid material.

The paper is organized as follows. For a compressible saturated porous media, Section 2 gives a brief introduction of the fundamentals of continuum description, such as the required kinematics, conservation equations and the multiplicative split of the deformation gradient. Section 3 provides the constitutive relations based on two sets of independent variables. The decoupled and coupled constitutive relations are developed and discussed in Section 4. In Section 5, the linearized version of the proposed constitutive relations is derived. Section 6 discusses the comparison of the linearized version of the proposed constitutive model with the models by Biot, by Lopatnikov and Cheng, and by Serpieri and Rosati. The definition of effective stress is outlined in section 7. In section 8, the proposed model is also compared with some existing experimental results.

## 2. Continuum mechanics for fluid-saturated porous media

### 2.1 Preliminaries

One of the most important concepts in porous media theory is the volume fraction. The volume fraction is defined as a statistical distribution of the individual constituents over the whole control space. Consider a fluid-saturated porous media, the volume fraction $n^\alpha$ can be formulated as

$$n^\alpha(\mathbf{x},t) = \frac{dV^\alpha}{dV}, \quad \alpha = \text{S, F} \tag{1}$$

where $\mathbf{x}$ is the actual position at time $t$; $dV$ is the bulk volume of the control space; $dV^\alpha$ is the real volume of the constituent $\alpha$, where $\alpha = \text{S, F}$ denotes the solid (S) and Fluid (F), respectively. The volume fractions satisfy

$$n^\text{S} + n^\text{F} = 1. \tag{2}$$

The density $\rho^{\alpha\text{R}}$ of the real material and the partial density $\rho^\alpha$ are quantified by the local mass $dm^\alpha$, the local volume $dV^\alpha$ of the real material and the bulk volume $dV$ as

$$\rho^{\alpha\text{R}}(\mathbf{x},t) = \frac{dm^\alpha}{dV^\alpha}, \quad \alpha = \text{S, F} \tag{3}$$

$$\rho^\alpha(\mathbf{x},t) = \frac{dm^\alpha}{dV}, \quad \alpha = \text{S, F} \tag{4}$$

where the superscript $\alpha R$ denotes the real material of the constituent $\alpha$.

The partial density and real density satisfy the relation

$$\rho^\alpha = n^\alpha \rho^{\alpha R}, \ \alpha = S, F. \tag{5}$$

As discussed by de Boer (2005) and Ehlers (2018), equations (3)-(5) are of crucial importance for the definition of the compressibility or incompressibility of material. The constituent $\alpha$ is considered incompressible when $\rho^{\alpha R}$ is taken as a constant. However, the partial density can vary through the changes of the volume fractions even if the constituent $\alpha$ is incompressible. Therefore, the volumetric deformation of compressible porous media should be described based on two variables, i.e., the real density $\rho^{\alpha R}$ and the volume fraction $n^\alpha$.

## 2.2 Basic kinematics

The kinematics in the porous media theory is based on the assumption that each spatial point $\mathbf{x}$ of the current configuration is simultaneously occupied by material particles of all constituents $\alpha$ (de Boer, 2005; Ehlers, 2002, 2009). The motion of these particles ($P^S$, $P^F$) should be considered as proceeding from different reference positions at time $t = t_0$, as shown in Fig. 1.

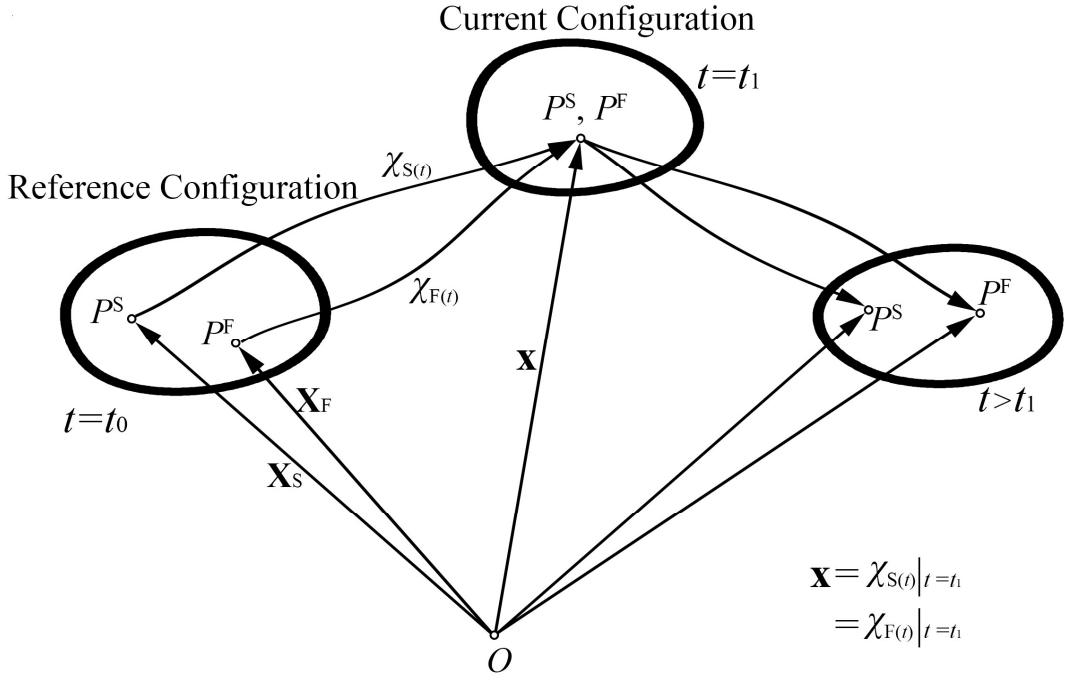

Fig.1 Kinematics of the porous media theory (redrawn from Ehlers, 2009).

The individual motion functions of solid and fluid read

$$\mathbf{x} = \chi_\alpha(\mathbf{X}_\alpha, t), \ \alpha = S, F. \tag{6}$$

The displacement vector is given by

$$\mathbf{u}_\alpha = \mathbf{x} - \mathbf{X}_\alpha, \ \alpha = S, F. \tag{7}$$

We introduce the deformation gradient of each phase as

$$\mathbf{F}_\alpha = \frac{\partial \chi_\alpha}{\partial \mathbf{X}_\alpha} = \mathrm{Grad}_\alpha \mathbf{x}, \ \alpha = \mathrm{S, F}. \tag{8}$$

The operator 'Grad$_\alpha$' denotes the gradient with respect to the reference position $\mathbf{X}_\alpha$ of constituent $\alpha$. The assumption of unique motion function requires that the Jacobian determinant of each phase is positive as

$$\mathbf{X}_\alpha = \chi_\alpha^{-1}(\mathbf{x},t), \ \alpha = \mathrm{S, F} \tag{9}$$

$$J_\alpha = \det \mathbf{F}_\alpha = \frac{dV^\alpha}{dV_0^\alpha} \neq 0, \ \alpha = \mathrm{S, F}. \tag{10}$$

The right and left Cauchy-Green deformation tensors $\mathbf{C}_\alpha$ and $\mathbf{b}_\alpha$, as well as the Green strain tensor $\mathbf{E}_\alpha$ are defined as

$$\mathbf{C}_\alpha = \mathbf{F}_\alpha^\mathrm{T} \mathbf{F}_\alpha, \ \alpha = \mathrm{S, F} \tag{11}$$

$$\mathbf{b}_\alpha = \mathbf{F}_\alpha \mathbf{F}_\alpha^\mathrm{T}, \ \alpha = \mathrm{S, F} \tag{12}$$

$$d\mathbf{x} \cdot d\mathbf{x} - d\mathbf{X}_\alpha \cdot d\mathbf{X}_\alpha = d\mathbf{X}_\alpha \cdot 2\mathbf{E}_\alpha \cdot d\mathbf{X}_\alpha, \ \alpha = \mathrm{S, F} \tag{13}$$

where the relation between $\mathbf{C}_\alpha$ and $\mathbf{E}_\alpha$ is

$$\mathbf{E}_\alpha = \frac{1}{2}(\mathbf{C}_\alpha - \mathbf{I}), \ \alpha = \mathrm{S, F}. \tag{14}$$

With the Lagrange description of the motion, the velocity and acceleration of each constituent are given by

$$\dot{\mathbf{x}}_\alpha = \frac{\partial \chi_\alpha(\mathbf{X}_\alpha, t)}{\partial t}, \ \alpha = \mathrm{S, F} \tag{15}$$

$$\ddot{\mathbf{x}}_\alpha = \frac{\partial^2 \chi_\alpha(\mathbf{X}_\alpha, t)}{\partial t^2}, \ \alpha = \mathrm{S, F}. \tag{16}$$

The Eulerian velocity and acceleration can be obtained by substitution of material coordinate $\mathbf{X}_\alpha$ with $\chi_\alpha^{-1}(\mathbf{x},t)$, i.e.

$$\mathbf{v}_\alpha = \dot{\mathbf{x}}_\alpha\left(\chi_\alpha^{-1}(\mathbf{x},t), t\right), \ \alpha = \mathrm{S, F} \tag{17}$$

$$\mathbf{a}_\alpha = \ddot{\mathbf{x}}_\alpha\left(\chi_\alpha^{-1}(\mathbf{x},t), t\right), \ \alpha = \mathrm{S, F}. \tag{18}$$

The spatial velocity gradient can be introduced as

$$\mathbf{L}_\alpha = \mathrm{grad}(\mathbf{v}_\alpha), \ \alpha = \mathrm{S, F}. \tag{19}$$

The operator 'grad' denotes the gradient referring to the spatial coordinate $\mathbf{X}$.

The symmetrical part, i.e. the rate of deformation tensor $\mathbf{D}_\alpha$, and the skew-symmetric part, i.e. the spin tensor $\mathbf{W}_\alpha$, of the velocity gradient $\mathbf{L}_\alpha$ are introduced as

$$\mathbf{D}_\alpha = \frac{1}{2}\left(\mathbf{L}_\alpha + \mathbf{L}_\alpha^{\mathrm{T}}\right), \ \alpha = \mathrm{S, F} \tag{20}$$

$$\mathbf{W}_\alpha = \frac{1}{2}\left(\mathbf{L}_\alpha - \mathbf{L}_\alpha^{\mathrm{T}}\right), \ \alpha = \mathrm{S, F} \tag{21}$$

where $\mathbf{L}_\alpha^{\mathrm{T}}$ is the transpose of $\mathbf{L}_\alpha$.

## 2.3. Conservation equations

The fundamental physical laws of porous media, such as conservation of mass, momentum and energy used in the present paper are briefly summarized in this section. According to the sign convention in rational mechanics, tension and compression are positive and negative, respectively.

In this paper, the conservation equations are given in the current configuration and described by the motions of each constituent $\alpha$ relative to the motion of the solid phase. It is convenient to denote the material derivative following the motion of the solid phase by $\mathrm{d}^{\mathrm{S}}(\cdot)/\mathrm{d}t$.

We consider a control volume $B_{\mathrm{S}}(t)$ fixed in the solid phase which is bounded by the surface $\partial B_{\mathrm{S}}(t)$. Let $\mathbf{n}$ be the unit normal at the surface $ds$ of the control volume $B_{\mathrm{S}}(t)$. It is worth to point out that the change of the control volume with time follows the motion of the solid phase.

A mass exchange between the different phases is not considered. The conservation law of mass for each phase requires that the rate of change of mass inside the control volume equals the mass flow rates in or out of the control volume. For the constituent $\alpha$ the conservation law of mass can be represented as

$$\frac{d^{\mathrm{S}}}{dt}\left(\int_{B_{\mathrm{S}}} \rho^\alpha dv\right) + \int_{\partial B_{\mathrm{S}}} \rho^\alpha \cdot (\mathbf{v}_{\alpha\mathrm{S}} \cdot \mathbf{n})ds = 0, \ \alpha = \mathrm{S, F} \tag{22}$$

where $\mathbf{v}_{\alpha\mathrm{S}}$ is the relative velocity of constituent $\alpha$, which is defined as the velocity relative to the motion of the solid phase, i.e.

$$\mathbf{v}_{\mathrm{SS}} = \mathbf{0} \tag{23}$$

$$\mathbf{v}_{\mathrm{FS}} = \mathbf{v}_{\mathrm{F}} - \mathbf{v}_{\mathrm{S}}. \tag{24}$$

With the divergence theorem, the conservation of mass of each phase becomes

$$\frac{d^{\mathrm{S}}(\rho^\alpha)}{dt} + \rho^\alpha \mathrm{div}(\mathbf{v}_{\mathrm{S}}) + \rho^\alpha \mathrm{div}(\mathbf{v}_{\alpha\mathrm{S}}) + (\mathbf{v}_{\alpha\mathrm{S}}) \cdot \mathrm{grad}(\rho^\alpha) = 0, \ \alpha = \mathrm{S, F}. \tag{25}$$

The conservation of momentum requires that the rate of change of the momentum is equal to the sum of external forces. It takes the form

$$\frac{d^{\mathrm{S}}}{dt}\left(\int_{B_{\mathrm{S}}} \rho^\alpha \mathbf{v}_\alpha dv\right) + \int_{\partial B_{\mathrm{S}}} \rho^\alpha \mathbf{v}_\alpha \cdot (\mathbf{v}_{\alpha\mathrm{S}} \cdot \mathbf{n})ds = \int_{B_{\mathrm{S}}} \rho^\alpha \mathbf{b}^\alpha dv + \int_{\partial B_{\mathrm{S}}} \mathbf{t}^\alpha ds + \int_{B_{\mathrm{S}}} \hat{\mathbf{p}}^\alpha dv, \ \alpha = \mathrm{S, F} \tag{26}$$

where $\rho^\alpha \mathbf{b}^\alpha$ and $\mathbf{t}^\alpha$ are the volume force and surface force of each phase, respectively. The interaction force $\hat{\mathbf{p}}^\alpha$ for each phase represents the drag on the relevant phase by the surrounding constituents.

By applying the divergence theorem, the conservation law of linear momentum is obtained as

$$\text{div}\mathbf{T}^\alpha + \rho^\alpha \mathbf{b}^\alpha + \hat{\mathbf{p}}^\alpha = \rho^\alpha \mathbf{a}_\alpha, \ \alpha = S, F \tag{27}$$

where $\mathbf{T}^\alpha$ is the partial Cauchy stress tensor which satisfies $\mathbf{t}^\alpha = \mathbf{T}^\alpha \mathbf{n}$ and $\mathbf{a}_\alpha$ is the acceleration. The total Cauchy stress tensor is the sum of the partial Cauchy stress tensors, i.e.

$$\mathbf{T} = \mathbf{T}^S + \mathbf{T}^F. \tag{28}$$

The interaction forces $\hat{\mathbf{p}}^S$ and $\hat{\mathbf{p}}^F$ satisfy the requirement

$$\hat{\mathbf{p}}^S + \hat{\mathbf{p}}^F = \mathbf{0}. \tag{29}$$

The conservation of energy states the sum of the rates of change of the internal and kinetic energies equal the rates of change of the mechanical work and the heat, i.e.

$$\underbrace{\frac{d^S}{dt}\left(\int_{B_S} \rho^\alpha \varepsilon^\alpha dv\right) + \int_{\partial B_S} \rho^\alpha \varepsilon^\alpha \cdot (\mathbf{v}_{\alpha S} \cdot \mathbf{n}) ds}_{\text{internal energy rate}} + \underbrace{\frac{d^S}{dt}\left(\int_{B_S} \frac{1}{2}\rho^\alpha \mathbf{v}_\alpha \cdot \mathbf{v}_\alpha dv\right) + \int_{\partial B_S} \frac{1}{2}\rho^\alpha \mathbf{v}_\alpha \cdot \mathbf{v}_\alpha \cdot (\mathbf{v}_{\alpha S} \cdot \mathbf{n}) ds}_{\text{kinetic energy rate}}$$
$$= \underbrace{\int_{B_S} \mathbf{v}_\alpha \cdot (\rho^\alpha \mathbf{b}^\alpha + \hat{\mathbf{p}}^\alpha) dv + \int_{\partial B_S} \mathbf{v}_\alpha \cdot \mathbf{t}^\alpha ds}_{\text{mechanical work}} + \underbrace{\int_{B_S} \rho^\alpha r^\alpha dv - \int_{\partial B_S} \mathbf{q}^\alpha \cdot \mathbf{n} ds}_{\text{heat}} + \underbrace{\int_{B_S} \hat{e}^\alpha dv}_{\text{energy transfer}}, \ \alpha = S, F \tag{30}$$

where $\varepsilon^\alpha$ is the specific internal energy, $r^\alpha$ the partial energy source, $\mathbf{q}^\alpha$ the partial heat flux vector, $\hat{e}^\alpha$ the energy transferred between two phases.

By using the divergence theorem, the conservation of energy can be rewritten as

$$\rho^\alpha \frac{d^S \varepsilon^\alpha}{dt} + \rho^\alpha \text{grad}(\varepsilon^\alpha) \cdot \mathbf{v}_{\alpha S} = \mathbf{T}^\alpha : \mathbf{L}_\alpha + \rho^\alpha r^\alpha - \text{div}(\mathbf{q}^\alpha) + \hat{e}^\alpha, \ \alpha = S, F. \tag{31}$$

~~It can be noted that~~ The sum of the energy transferred between all of the phases and work done by the interaction forces must be zero, i.e.

$$\sum_{\alpha = S, F} (\mathbf{v}_\alpha \cdot \hat{\mathbf{p}}^\alpha + \hat{e}^\alpha) = 0. \tag{32}$$

## 2.4 Multiplicative split of the deformation gradient

For further investigations, it is convenient to introduce the multiplicative split of the deformation gradient of the solid phase, i.e. the solid skeleton, $\mathbf{F}_S$, into its purely volumetric part, $\overline{\mathbf{F}}_S$, and the volume-preserving part, $\breve{\mathbf{F}}_S$ (Simo, 1988; Bluhm and de Boer, 1997), i.e.

$$\mathbf{F}_S = \overline{\mathbf{F}}_S \breve{\mathbf{F}}_S. \tag{33}$$

By using this concept of multiplicative split, we obtain

$$\overline{\mathbf{F}}_S = (J_S)^{1/3} \mathbf{I} \tag{34}$$

$$\mathbf{C}_S = (J_S)^{2/3} \breve{\mathbf{C}}_S \tag{35}$$

$$\breve{\mathbf{C}}_S = \breve{\mathbf{F}}_S^T \breve{\mathbf{F}}_S \tag{36}$$

$$\mathbf{b}_S = (J_S)^{2/3} \breve{\mathbf{b}}_S \tag{37}$$

$$\breve{\mathbf{b}}_S = \breve{\mathbf{F}}_S \breve{\mathbf{F}}_S^T \tag{38}$$

where $\breve{\mathbf{C}}_S$ and $\breve{\mathbf{b}}_S$ are, respectively, the volume-preserving part of the right and left Cauchy deformation tensor and $\breve{\mathbf{F}}_S^T$ is the transpose of $\breve{\mathbf{F}}_S$. The material time-derivative of $J_S$ and $\breve{\mathbf{C}}_S$ yields (de Boer, 2005)

$$\frac{d^S J_S}{dt} = J_S (\mathbf{D}_S : \mathbf{I}) \tag{39}$$

$$\frac{d^S \breve{\mathbf{C}}_S}{dt} = 2 J_S^{-2/3} \mathbf{F}_S^T \mathbf{D}_S^D \mathbf{F}_S \tag{40}$$

where $\mathbf{D}_S^D$ is the deviatoric part of $\mathbf{D}_S$.

Further, the volumetric part of the deformation gradient for the solid skeleton, $\overline{\mathbf{F}}_S$, can be multiplicatively decomposed into two parts as (Bluhm and de Boer, 1997; Ehlers, 2018)

$$\overline{\mathbf{F}}_S = \overline{\mathbf{F}}_{SN} \overline{\mathbf{F}}_{SR} \tag{41}$$

where $\overline{\mathbf{F}}_{SN}$ is the part of $\overline{\mathbf{F}}_S$ which governs the change of volume fraction of solid phase denoted by subscript SN, and $\overline{\mathbf{F}}_{SR}$ is the part of $\overline{\mathbf{F}}_S$ governing the volumetric deformation of real solid material denoted by subscript SR. By using the mass conservation equation (23), the following relations can be obtained

$$J_S = \det \mathbf{F}_S = \frac{\rho_0^S}{\rho^S} \tag{42}$$

$$J_{SN} = \det \overline{\mathbf{F}}_{SN} = \frac{n_0^S}{n^S} \tag{43}$$

$$J_{SR} = \det \overline{\mathbf{F}}_{SR} = \frac{\rho_0^{SR}}{\rho^{SR}}. \tag{44}$$

The multiplicative decomposition of equation (41) establishes the relationship between the macroscopic deformation gradient $\overline{\mathbf{F}}_S$ and the microscopic deformation gradients $\overline{\mathbf{F}}_{SN}$ and $\overline{\mathbf{F}}_{SR}$. It provides the theoretical basis for the investigation of the macroscopic constitutive relations by using the independent variables $n^S$ and $\rho^{SR}$ which are introduced at the microscopic scale. There are several other approaches to the multiplicative split of the deformation gradient for the solid skeleton as discussed for instance by Gajo (2010) and by Hu (2018).

# 3. Constitutive relations of compressible saturated porous media based on two sets of independent variables

In order to close the model, the constitutive relations for the compressible fluid-saturated porous media are required. Isotropic and elastic material properties are considered for each constituent. In general, thermodynamic admissibility conditions for processes can be derived from the Clausius-Duhem inequality by means of Coleman-Noll method (Coleman and Noll, 1963) or Müller-Liu method (Müller, 1985). The Coleman-Noll argument has made substantial contributions to the field of modeling the constitutive relations of porous media (de Boer, 2000, 2005; Ehlers, 2002, 2009, 2018). On the other hand, the Muller-Liu method has also been applied in the constitutive modeling of porous media (Wilmanski, 2008; Liu, 2014). In this section, the Clausius-Duhem inequality for the compressible saturated porous media is introduced based on the control volume fixed in the solid phase and used for establishing restrictions of constitutive relations to physically acceptable processes by using the Coleman-Noll arguments.

## 3.1 Clausius-Duhem Inequality

The overall Clausius-Duhem inequality for all constituents is described as (de Boer, 2005)

$$\sum_{\alpha=S,F}\left[\frac{d^S}{dt}\left(\int_{B_S}\rho^\alpha\eta^\alpha dv\right)+\int_{\partial B_S}\rho^\alpha\eta^\alpha\cdot(\mathbf{v}_{\alpha S}\cdot\mathbf{n})ds\right]\geq\sum_{\alpha=S,F}\left[\int_{B_S}\frac{1}{\theta^\alpha}\rho^\alpha r^\alpha dv-\int_{\partial B_S}\frac{1}{\theta^\alpha}\mathbf{q}^\alpha\cdot\mathbf{n}ds\right] \quad (45)$$

where $\eta^\alpha$ is the specific entropy and $\theta^\alpha$ is the temperature of constituent $\alpha$. The specific Helmholtz free energy can be written as

$$\psi^\alpha=\varepsilon^\alpha-\theta^\alpha\eta^\alpha,\ \alpha=S,F. \quad (46)$$

Substituting the conservation equation of energy (31), Helmholtz free energy expression (46) and relation (32) into the Clausius-Duhem inequality (45) yields

$$\sum_{\alpha=S,F}\left\{\begin{array}{l}-\rho^\alpha\left[\eta^\alpha\dfrac{d^S\theta^\alpha}{dt}+\eta^\alpha\mathrm{grad}(\theta^\alpha)\cdot\mathbf{v}_{\alpha S}+\dfrac{d^S\psi^\alpha}{dt}+\mathrm{grad}(\psi^\alpha)\cdot\mathbf{v}_{\alpha S}\right]\\ +\mathbf{T}^\alpha:\mathbf{L}_\alpha-\mathbf{v}_\alpha\cdot\hat{\mathbf{p}}^\alpha-\dfrac{1}{\theta^\alpha}\mathbf{q}^\alpha\cdot\mathrm{grad}(\theta^\alpha)\end{array}\right\}\geq 0. \quad (47)$$

Since $\mathbf{T}^\alpha$ is symmetric, the following relation is introduced

$$\mathbf{T}^\alpha:\mathbf{L}_\alpha=\mathbf{T}^\alpha:\mathbf{D}_\alpha,\ \alpha=S,F \quad (48)$$

For the sake of simplicity, we assume same and constant temperatures of the two phases. The internal viscous shear stresses in the fluid are assumed negligible. Based on the volumetric-deviatoric multiplicative split (Holzapfel, 2000), a decoupled work input part of solid phase can be introduced as

$$\mathbf{T}^S : \mathbf{D}_S = \breve{\mathbf{T}}^S : \mathbf{D}_S^D - p^S \frac{d^S \rho^S}{\rho^S} \tag{49}$$

where $\breve{\mathbf{T}}^S$ is the partial fictitious Cauchy stress tensor and $p^S$ is defined as

$$p^S = \frac{1}{3} \mathbf{T}^S : \mathbf{I}. \tag{50}$$

The mechanical work part of fluid phase can be written as

$$\mathbf{T}^F : \mathbf{L}_F = -p^F \frac{d^S \rho^F}{\rho^F} - \frac{p^F}{\rho^F} \operatorname{grad}(\rho^F) \cdot \mathbf{v}_{FS} \tag{51}$$

where $p^F$ is introduced as

$$p^F = \frac{1}{3} \mathbf{T}^F : \mathbf{I}. \tag{52}$$

With the substitution of equations (49), (50) and (52) into (47), the Clausius-Duhem inequality reduces to

$$\begin{aligned} &-\rho^S \frac{d^S \psi^S}{dt} - \rho^F \frac{d^S \psi^F}{dt} - \rho^F \operatorname{grad}(\psi^F) \cdot \mathbf{v}_{FS} + \breve{\mathbf{T}}^S : \mathbf{D}_S^D \\ &-p^S \frac{1}{\rho^S} \frac{d^S \rho^S}{dt} - p^F \frac{1}{\rho^F} \frac{d^S \rho^F}{dt} - \frac{p^F}{\rho^F} \operatorname{grad}(\rho^F) \cdot \mathbf{v}_{FS} - \hat{\mathbf{p}}^F \cdot \mathbf{v}_{FS} \geq 0 \end{aligned} \tag{53}$$

The following relations are obtained by the time derivative of the partial density in equation (5)

$$\frac{1}{\rho^S} \frac{d^S \rho^S}{dt} = \frac{1}{n^S} \frac{d^S n^S}{dt} + \frac{1}{\rho^{SR}} \frac{d^S \rho^{SR}}{dt} \tag{54}$$

$$\frac{1}{\rho^F} \frac{d^S \rho^F}{dt} = \frac{1}{n^F} \frac{d^S n^F}{dt} + \frac{1}{\rho^{FR}} \frac{d^S \rho^{FR}}{dt} \tag{55}$$

where $\dfrac{d^S n^S}{dt}$ and $\dfrac{d^S n^F}{dt}$ satisfy

$$\frac{d^S n^S}{dt} + \frac{d^S n^F}{dt} = 0. \tag{56}$$

In this paper, two sets of independent variables are considered for the Helmholtz free energy function

$$\text{set (a):} \quad \{\breve{\mathbf{C}}_S, n^S, \rho^{SR}, \rho^{FR}\} \tag{57}$$

$$\text{set (b):} \quad \{\breve{\mathbf{C}}_S, \rho^S, \rho^{SR}, \rho^{FR}\}. \tag{58}$$

In the following it is shown that the constitutive relations for the compressible fluid-saturated porous media can be derived by combining the results based on the two sets (a) and (b). The results obtained based on set (a) are related to the decoupled form of free energy which is convenient for simplifying the constitutive relation. However, it is difficult to use because both $n^S$ and $\rho^{SR}$ are introduced at the microscopic scale. For this reason, the partial density of the solid phase $\rho^S$, where $\rho^S = n^S \rho^{SR}$, is introduced as an alternative independent variable in set (b).

The principle of phase separation (Passman et al., 1984; Ehlers, 2018), i.e., the free energy of each constituent depends only on its own constitutive variables on its microscope, provides the basis for the present research. Following the approach used by Ehlers (2018), the free energy functions of the solid and fluid phases on set (a) is introduced as

$$\psi^S = \tilde{\psi}^S \left( \breve{\mathbf{C}}_S, n^S, \rho^{SR} \right) \tag{59}$$

$$\psi^F = \tilde{\psi}^F \left( \rho^{FR} \right). \tag{60}$$

The free energy functions can also be expressed on set (b) as

$$\psi^S = \hat{\psi}^S \left( \breve{\mathbf{C}}_S, \rho^S, \rho^{SR} \right) \tag{61}$$

$$\psi^F = \hat{\psi}^F \left( \rho^{FR} \right). \tag{62}$$

These free energy functions satisfy

$$\psi^S = \tilde{\psi}^S \left( \breve{\mathbf{C}}_S, n^S, \rho^{SR} \right) = \hat{\psi}^S \left( \breve{\mathbf{C}}_S, \rho^S, \rho^{SR} \right) \tag{63}$$

$$\psi^F = \tilde{\psi}^F \left( \rho^{FR} \right) = \hat{\psi}^F \left( \rho^{FR} \right). \tag{64}$$

## 3.2 Constitutive relations based on the independent variables of set (a)

The material derivative of the Helmholtz free energy with respect to the motion of solid phase can be obtained based on set (a) as

$$\frac{d^S \tilde{\psi}^S}{dt} = \frac{\partial \tilde{\psi}^S}{\partial \breve{\mathbf{C}}_S} : \frac{d^S \breve{\mathbf{C}}_S}{dt} + \frac{\partial \tilde{\psi}^S}{\partial n^S} \frac{d^S n^S}{dt} + \frac{\partial \tilde{\psi}^S}{\partial \rho^{SR}} \frac{d^S \rho^{SR}}{dt} \tag{65}$$

$$\frac{d^S \tilde{\psi}^F}{dt} = \frac{\partial \tilde{\psi}^F}{\partial \rho^{FR}} \frac{d^S \rho^{FR}}{dt}. \tag{66}$$

The spatial gradient of free energy reads

$$\operatorname{grad} \tilde{\psi}^F = \frac{\partial \tilde{\psi}^F}{\partial \rho^{FR}} \operatorname{grad} \left( \rho^{FR} \right). \tag{67}$$

Substituting (40), (54)-(56) and (65)-(67) into (53) yields

$$\left(\tilde{\mathbf{T}}^{\mathrm{S}}-2\rho^{\mathrm{S}}J_{\mathrm{S}}^{-2/3}\mathbf{F}_{\mathrm{S}}\frac{\partial\tilde{\psi}^{\mathrm{S}}}{\partial\breve{\mathbf{C}}_{\mathrm{S}}}\mathbf{F}_{\mathrm{S}}^{\mathrm{T}}\right):\mathbf{D}_{\mathrm{S}}^{\mathrm{D}}-\left(P-\frac{p^{\mathrm{F}}}{n^{\mathrm{F}}}+n^{\mathrm{S}}\rho^{\mathrm{S}}\frac{\partial\tilde{\psi}^{\mathrm{S}}}{\partial n^{\mathrm{S}}}\right)\frac{1}{n^{\mathrm{S}}}\frac{d^{\mathrm{S}}n^{\mathrm{S}}}{dt}$$

$$-\left(p^{\mathrm{S}}+\rho^{\mathrm{S}}\rho^{\mathrm{SR}}\frac{\partial\tilde{\psi}^{\mathrm{S}}}{\partial\rho^{\mathrm{SR}}}\right)\frac{1}{\rho^{\mathrm{SR}}}\frac{d^{\mathrm{S}}\rho^{\mathrm{SR}}}{dt}-\left(p^{\mathrm{F}}+\rho^{\mathrm{F}}\rho^{\mathrm{FR}}\frac{\partial\tilde{\psi}^{\mathrm{F}}}{\partial\rho^{\mathrm{FR}}}\right)\frac{1}{\rho^{\mathrm{FR}}}\frac{d^{\mathrm{S}}\rho^{\mathrm{FR}}}{dt} \qquad (68)$$

$$-\left[\hat{\mathbf{p}}^{\mathrm{F}}+\frac{p^{\mathrm{F}}}{\rho^{\mathrm{F}}}\mathrm{grad}\left(\rho^{\mathrm{F}}\right)+\rho^{\mathrm{F}}\frac{\partial\tilde{\psi}^{\mathrm{F}}}{\partial\rho^{\mathrm{FR}}}\mathrm{grad}\left(\rho^{\mathrm{FR}}\right)\right]\cdot\mathbf{v}_{\mathrm{FS}}\geq 0$$

in which

$$\frac{\partial\tilde{\psi}^{\mathrm{S}}}{\partial\breve{\mathbf{C}}_{\mathrm{S}}}:\frac{d^{\mathrm{S}}\breve{\mathbf{C}}_{\mathrm{S}}}{dt}=\frac{\partial\tilde{\psi}^{\mathrm{S}}}{\partial\breve{\mathbf{C}}_{\mathrm{S}}}:2J_{\mathrm{S}}^{-2/3}\mathbf{F}_{\mathrm{S}}^{\mathrm{T}}\mathbf{D}_{\mathrm{S}}^{\mathrm{D}}\mathbf{F}_{\mathrm{S}}=2J_{\mathrm{S}}^{-2/3}\mathbf{F}_{\mathrm{S}}\frac{\partial\tilde{\psi}^{\mathrm{S}}}{\partial\breve{\mathbf{C}}_{\mathrm{S}}}\mathbf{F}_{\mathrm{S}}^{\mathrm{T}}:\mathbf{D}_{\mathrm{S}}^{\mathrm{D}} \qquad (69)$$

is considered and the total mean pressure $P$ is the sum of the mean pressure $p^{\mathrm{S}}$ of the solid phase and the mean pressure $p^{\mathrm{F}}$ of the fluid phase, i.e.

$$P=\frac{1}{3}(\mathbf{T}:\mathbf{I})=p^{\mathrm{S}}+p^{\mathrm{F}}. \qquad (70)$$

The Coleman-Noll argument states that the Clausius-Duhem inequality must be true for any value of the independent variables and any possible physical processes. Applying Coleman-Noll argument yields the result that in the inequality (68) the coefficients of the terms of the time rates of independent variables should be vanish, i.e. for the part of the

solid phase:
$$\tilde{\mathbf{T}}^{\mathrm{S}}=2\rho^{\mathrm{S}}J_{\mathrm{S}}^{-2/3}\mathbf{F}_{\mathrm{S}}\frac{\partial\tilde{\psi}^{\mathrm{S}}}{\partial\breve{\mathbf{C}}_{\mathrm{S}}}\mathbf{F}_{\mathrm{S}}^{\mathrm{T}} \qquad (71)$$

$$P-\frac{p^{\mathrm{F}}}{n^{\mathrm{F}}}=-n^{\mathrm{S}}\rho^{\mathrm{S}}\frac{\partial\tilde{\psi}^{\mathrm{S}}}{\partial n^{\mathrm{S}}} \qquad (72)$$

$$p^{\mathrm{S}}=-\rho^{\mathrm{S}}\rho^{\mathrm{SR}}\frac{\partial\tilde{\psi}^{\mathrm{S}}}{\partial\rho^{\mathrm{SR}}} \qquad (73)$$

and fluid phase:
$$p^{\mathrm{F}}=-\rho^{\mathrm{F}}\rho^{\mathrm{FR}}\frac{\partial\tilde{\psi}^{\mathrm{F}}}{\partial\rho^{\mathrm{FR}}} \qquad (74)$$

Equations (71)-(74) are the constitutive equations based on the variables of set (a). With the substitution of equations (5) and (74) into (68), the dissipation part of the inequality may be reduced to

$$-\left[\hat{\mathbf{p}}^{\mathrm{F}}+\frac{p^{\mathrm{F}}}{\rho^{\mathrm{F}}}\mathrm{grad}\left(\rho^{\mathrm{F}}\right)+\rho^{\mathrm{F}}\frac{\partial\tilde{\psi}^{\mathrm{F}}}{\partial\rho^{\mathrm{FR}}}\mathrm{grad}\left(\rho^{\mathrm{FR}}\right)\right]\cdot\mathbf{v}_{\mathrm{FS}}\geq 0. \qquad (75)$$

## 3.3 Constitutive relations based on the independent variables of set (b)

In order to obtain the constitutive relations based on the independent variables of set (b) the Coleman-Noll argument is applied. With respect to the motion of solid phase the material derivative of the free energy can be obtained as

$$\frac{d^S\hat{\psi}^S}{dt} = \frac{\partial \hat{\psi}^S}{\partial \breve{\mathbf{C}}_S} : \frac{d^S \breve{\mathbf{C}}_S}{dt} + \frac{\partial \hat{\psi}^S}{\partial \rho^S} \frac{d^S \rho^S}{dt} + \frac{\partial \hat{\psi}^S}{\partial \rho^{SR}} \frac{d^S \rho^{SR}}{dt} \tag{76}$$

$$\frac{d^S\hat{\psi}^F}{dt} = \frac{\partial \hat{\psi}^F}{\partial \rho^{FR}} \frac{d^S \rho^{FR}}{dt}. \tag{77}$$

The spatial gradient of the free energy reads

$$\mathrm{grad}\hat{\psi}^F = \frac{\partial \hat{\psi}^F}{\partial \rho^{FR}} \mathrm{grad}\rho^{FR}. \tag{78}$$

With the substitution of (40), (54)-(56) and (76)-(78) into (53), the Clausius-Duhem inequality can be expressed as

$$\begin{aligned}
&\left(\tilde{\mathbf{T}}^S - 2\rho^S J_S^{-2/3} \mathbf{F}_S \frac{\partial \hat{\psi}^S}{\partial \breve{\mathbf{C}}_S} \mathbf{F}_S^T \right) : \mathbf{D}_S^D - \left[ P - \frac{p^F}{n^F} + \left(\rho^S\right)^2 \frac{\partial \hat{\psi}^S}{\partial \rho^S} \right] \frac{1}{\rho^S} \frac{d^S \rho^S}{dt} \\
&- \left(\frac{n^S}{n^F} p^F + \rho^S \rho^{SR} \frac{\partial \hat{\psi}^S}{\partial \rho^{SR}} \right) \frac{1}{\rho^{SR}} \frac{d^S \rho^{SR}}{dt} - \left(p^F + \rho^F \rho^{FR} \frac{\partial \hat{\psi}^F}{\partial \rho^{FR}} \right) \frac{1}{\rho^{FR}} \frac{d^S \rho^{FR}}{dt} \\
&- \left[\hat{\mathbf{p}}^F + \frac{p^F}{\rho^F} \mathrm{grad}\left(\rho^F\right) + \rho^F \frac{\partial \hat{\psi}^F}{\partial \rho^{FR}} \mathrm{grad}\left(\rho^{FR}\right) \right] \cdot \mathbf{v}_{FS} \geq 0
\end{aligned} \tag{79}$$

By applying an analogous procedure to Sect. 3.2, the constitutive relations obtained based on set (b) read

for the solid phase:
$$\tilde{\mathbf{T}}^S = 2\rho^S J_S^{-2/3} \mathbf{F}_S \frac{\partial \hat{\psi}^S}{\partial \breve{\mathbf{C}}_S} \mathbf{F}_S^T \tag{80}$$

$$P - \frac{p^F}{n^F} = -\left(\rho^S\right)^2 \frac{\partial \hat{\psi}^S}{\partial \rho^S} \tag{81}$$

$$\frac{n^S}{n^F} p^F = -\rho^S \rho^{SR} \frac{\partial \hat{\psi}^S}{\partial \rho^{SR}} \tag{82}$$

and for the liquid phase:
$$p^F = -\rho^F \rho^{FR} \frac{\partial \hat{\psi}^F}{\partial \rho^{FR}}. \tag{83}$$

The dissipation part of the constitutive relations based on set (b) is the same as that on set (a) as shown in equation (75). Several results similar to the constitutive relations (80)-(83) have been represented by Serpieri and Rosati (2011) based on the variational macroscopic theory of porous media and by Hu (2018) based on the pure macroscopic theory.

## 4. Decoupled and coupled constitutive relations

In this section, the decoupled constitutive relations based on set (a) is first considered. It is convenient to assume that the free energy can be decoupled into volumetric and volume-preserved parts based on the multiplicative split of deformation gradient. The decoupled free energy method dates back to Flory (1961) and has been widely adopted in various fields (Simo, 1988; Mosler and Bruhns, 2009; Drass et al., 2018). The decoupled form of the free energy is convenient for simplifying the constitutive relation, however, these constitutive relations are difficult to use in a direct manner in practical applications. For this reason the coupled constitutive relations are introduced secondarily based on set (b), which is convenient for the application.

## 4.1 Decoupled constitutive relations based on set (a)

The application of the deviatoric/volumetric split of the deformation gradient (33) to the free energy function (59) of set (a) yields

$$\tilde{\psi}^S\left(\breve{\mathbf{C}}_S, n^S, \rho^{SR}\right) = \tilde{\psi}_1^S\left(\breve{\mathbf{C}}_S\right) + \tilde{\psi}_2^S\left(n^S, \rho^{SR}\right) \tag{84}$$

where $\tilde{\psi}_1^S\left(\breve{\mathbf{C}}_S\right)$ denotes the volume-preserved part and $\tilde{\psi}_2^S\left(n^S, \rho^{SR}\right)$ is the volumetric part of the free energy of the solid phase. Based on the multiplicative split of the volumetric deformation gradient (41), it is convenient to assume that the free energy $\tilde{\psi}_2^S\left(n^S, \rho^{SR}\right)$ can be decoupled into two parts as

$$\tilde{\psi}_2^S\left(n^S, \rho^{SR}\right) = \tilde{\psi}_3^S\left(n^S\right) + \tilde{\psi}_4^S\left(\rho^{SR}\right) \tag{85}$$

where $\tilde{\psi}_3^S\left(n^S\right)$ and $\tilde{\psi}_4^S\left(\rho^{SR}\right)$ are the volume fraction part and the real solid part of the volumetric part of the free energy $\tilde{\psi}_2^S\left(n^S, \rho^{SR}\right)$, respectively. The decoupled form of free energy $\tilde{\psi}_2^S\left(n^S, \rho^{SR}\right)$ has been interpreted as "micro-homogenous" by Lopatnikov and Cheng (2002) or as "homogenized response" by Hu (2016). In the present paper, however, the decoupled behavior of free energy $\tilde{\psi}_2^S\left(n^S, \rho^{SR}\right)$ is introduced for the convenience in describing the compressibility of the real solid material. Following the multiplicative split of the free energy, the constitutive relations of the solid phase (71)-(73) may be rewritten as

$$\breve{\mathbf{T}}^S = 2\rho^S J_S^{-2/3} \mathbf{F}_S \frac{\partial \tilde{\psi}_1^S\left(\breve{\mathbf{C}}_S\right)}{\partial \breve{\mathbf{C}}_S} \mathbf{F}_S^T \tag{86}$$

$$P - \frac{p^F}{n^F} = -n^S \rho^S \frac{\partial \tilde{\psi}_3^S\left(n^S\right)}{\partial n^S} \tag{87}$$

$$p^S = -\rho^S \rho^{SR} \frac{\partial \tilde{\psi}_4^S\left(\rho^{SR}\right)}{\partial \rho^{SR}}. \tag{88}$$

It can be noted that several similar results to constitutive relations (86)-(88) have also been presented by Hu (2018) based on the pure macroscopic theory.

The deviatoric part of the partial stress of the solid phase, $\mathbf{T}^{SD}$, can be obtained by the partial fictitious Cauchy stress as (Kelly, 2020)

$$\mathbf{T}^{SD} = \left(\overset{4}{\mathbf{I}} - \frac{1}{3}\mathbf{I} \otimes \mathbf{I}\right) : \breve{\mathbf{T}}^S = 2\rho^S \mathbf{F}_S \frac{\partial \tilde{\psi}_1^S\left(\breve{\mathbf{C}}_S\right)}{\partial \mathbf{C}_S} \mathbf{F}_S^T. \tag{89}$$

In the following the decoupled constitutive relations related to the independent variables $\rho^{SR}$ and $\rho^{FR}$ will be discussed at the microscopic level. The free energy function, which is only related to the density of real solid material per unit volume, i.e. $\rho^S \tilde{\psi}_4^S (\rho^{SR})$, can also be rewritten as

$$\rho^S \tilde{\psi}_4^S (\rho^{SR}) = n^S \rho^{SR} \psi^{SR} (\rho^{SR}) \tag{90}$$

where $\psi^{SR}(\rho^{SR})$ is the free energy associated with the volumetric deformation per unit mass of the real solid material. Due to the relation $\rho^S = n^S \rho^{SR}$, the following relation holds

$$\tilde{\psi}_4^S (\rho^{SR}) = \psi^{SR} (\rho^{SR}). \tag{91}$$

Substituting equation (91) into (88) yields

$$p^S = -n^S (\rho^{SR})^2 \frac{\partial \psi^{SR} (\rho^{SR})}{\partial \rho^{SR}}. \tag{92}$$

The hydrostatic pressure of real solid material (Bowen, 1989; Ehlers, 2018) is introduced as

$$p^{SR} = (\rho^{SR})^2 \frac{\partial \psi^{SR} (\rho^{SR})}{\partial \rho^{SR}}. \tag{93}$$

From (92) and (93), we have

$$p^S = -n^S p^{SR} \tag{94}$$

It is worth noting that the specific free energy $\tilde{\psi}^F (\rho^{FR})$ is only related to the density of the real fluid material. The specific free energy of the fluid phase per unit volume, i.e. $\rho^F \tilde{\psi}_2^F (\rho^{FR})$, can be rewritten as

$$\rho^F \tilde{\psi}_2^F (\rho^{FR}) = n^F \rho^{FR} \psi^{FR} (\rho^{FR}) \tag{95}$$

where $\psi^{FR}(\rho^{FR})$ is the free energy associated with the volumetric deformation per unit mass of the real fluid material. Noting $\rho^F = n^F \rho^{FR}$, equation (95) reduces to

$$\tilde{\psi}^F (\rho^{FR}) = \psi^{FR} (\rho^{FR}). \tag{96}$$

Combining (5), (74) and (96) yields

$$p^F = -\rho^F \rho^{FR} \frac{\partial \psi^{FR} (\rho^{FR})}{\partial \rho^{FR}}. \tag{97}$$

The hydrostatic pressure of the real fluid material can also be introduced as

$$p^{FR} = \left(\rho^{FR}\right)^2 \frac{\partial \psi^{FR}\left(\rho^{FR}\right)}{\partial \rho^{FR}}. \tag{98}$$

A comparison of (97) and (98) yields

$$p^F = -n^F p^{FR}. \tag{99}$$

It is worth remarking that compared to the partial pressures $p^S$ and $p^F$ the hydrostatic pressure of the real solid material ($p^{SR}$) and the real fluid material ($p^{FR}$) may be more appropriate for dealing with engineering problems in practice. The constitutive relations (93) and (98) are also referred to as the intrinsic constitutive laws governing the real bulk stiffnesses of the solid and the fluid, respectively (Borja and Koliji, 2009).

It follows that the constitutive relation (87) may be rewritten as

$$P + p^{FR} = -n^S \rho^S \frac{\partial \tilde{\psi}_3^S\left(n^S\right)}{\partial n^S}. \tag{100}$$

Substitution of equation (27) together with (99) into (75) gives

$$-\left[n^F \mathrm{grad}\left(p^{FR}\right) - \rho^F \mathbf{b}^F + \rho^F \mathbf{a}_F\right] \cdot \mathbf{v}_{FS} \geq 0. \tag{101}$$

According to the Coleman-Noll argument, it requires

$$-n^F \mathbf{v}_{FS} \propto \mathrm{grad}\left(p^{FR}\right) - \rho^{FR} \mathbf{b}^F + \rho^{FR} \mathbf{a}_F. \tag{102}$$

As was pointed out by Coussy (2004), the simplest version of equation (102) is Darcy's law, i.e.

$$n^F \mathbf{v}_{FS} = -k\left[\mathrm{grad}\left(p^{FR}\right) - \rho^{FR} \mathbf{b}^F + \rho^{FR} \mathbf{a}_F\right] \tag{103}$$

where $k$ is the permeability of the fluid.

Equations (89), (93), (98) and (100) are the decoupled constitutive equations obtained based on set (a). Although simple, the decoupled equations (93) and (100) are difficult to use in practical applications due to the independent variables $n^S$ and $\rho^{SR}$, which are introduced at the microscopic scale. In order to overcome this disadvantage, the coupled constitutive relations will be considered in section 4.2.

### 4.2 Coupled constitutive relations based on set (b)

For the constitutive equations on set (b), the free energy of the solid phase may be split into two parts using the deviatoric/volumetric multiplicative split as

$$\hat{\psi}^S\left(\bar{\mathbf{C}}^S, \rho^S, \rho^{SR}\right) = \hat{\psi}_1^S\left(\bar{\mathbf{C}}^S\right) + \hat{\psi}_2^S\left(\rho^S, \rho^{SR}\right) \tag{104}$$

where the volume preserved part $\hat{\psi}_1^S\left(\bar{\mathbf{C}}^S\right)$ and the volumetric part $\hat{\psi}_2^S\left(\rho^S, \rho^{SR}\right)$ satisfies

$$\hat{\psi}_1^S\left(\bar{\mathbf{C}}^S\right) = \tilde{\psi}_1^S\left(\bar{\mathbf{C}}^S\right) \tag{105}$$

$$\hat{\psi}_2^S\left(\rho^S, \rho^{SR}\right) = \tilde{\psi}_3^S\left(n^S\right) + \tilde{\psi}_4^S\left(\rho^{SR}\right) = \tilde{\psi}_3^S\left(n^S\right) + \psi^{SR}\left(\rho^{SR}\right). \tag{106}$$

Substituting the above equations into (80)-(82) leads to the coupled constitutive relations for the solid phase. The overall constitutive relations based on set (b) can be summarized as

$$\mathbf{T}^{SD} = 2\rho^S \mathbf{F}_S \frac{\partial \tilde{\psi}_1^S\left(\breve{\mathbf{C}}_S\right)}{\partial \mathbf{C}_S} \mathbf{F}_S^T \tag{107}$$

$$P + p^{FR} = -\left(\rho^S\right)^2 \frac{\partial\left[\tilde{\psi}_3^S\left(n^S\right) + \psi^{SR}\left(\rho^{SR}\right)\right]}{\partial \rho^S} \tag{108}$$

$$n^S p^{FR} = \rho^S \rho^{SR} \frac{\partial\left[\tilde{\psi}_3^S\left(n^S\right) + \psi^{SR}\left(\rho^{SR}\right)\right]}{\partial \rho^{SR}} \tag{109}$$

$$p^{FR} = \left(\rho^{FR}\right)^2 \frac{\partial \psi^{FR}\left(\rho^{FR}\right)}{\partial \rho^{FR}}. \tag{110}$$

The proposed constitutive relations (108) and (109) couples the intrinsic constitutive model of the real solid material and the macroscopic constitutive relation of solid skeleton by using the decoupled free energy. As $\rho^S$ and $\rho^{SR}$ are independent variables in the constitutive relations, together with the relations (54) and (56), the evolution of porosity can also be obtained.

The constitutive relations for the solid phase (107)-(110) contain four free energy terms, i.e. $\tilde{\psi}_1^S\left(\breve{\mathbf{C}}_S\right)$, $\tilde{\psi}_3^S\left(n^S\right)$, $\psi^{SR}\left(\rho^{SR}\right)$ and $\psi^{FR}\left(\rho^{FR}\right)$, which are decoupled. Each of them is only related to a single independent variable. Instead of defining a free energy function $\hat{\psi}_2^S\left(\rho^S, \rho^{SR}\right)$ as shown in Eq. (106), it is more convenient for practical application to use the decoupled free energy functions $\tilde{\psi}_3^S\left(n^S\right)$ and $\psi^{SR}\left(\rho^{SR}\right)$. Therefore, equations (107)-(110) are the macro-microscopic coupled constitutive relations we obtained for the compressible fluid-saturated porous media.

# 5 Linearization of constitutive relations

In this section, the linearized version of the constitutive relations appropriate to situations in which the displacement gradient is small will be presented. The infinitesimal increment of the strain-like variables related to the independent variables may be introduced as

$$d\boldsymbol{\varepsilon}^S = \frac{1}{2}\left[\frac{\partial d\mathbf{u}_S}{\partial \mathbf{x}} + \left(\frac{\partial d\mathbf{u}_S}{\partial \mathbf{x}}\right)^T\right] \tag{111}$$

$$d\varepsilon_V^{SN} = \frac{dJ_{SN}}{J_{SN}} \tag{112}$$

$$d\varepsilon_{\text{V}}^{\text{SR}} = \frac{dJ_{\text{SR}}}{J_{\text{SR}}} \tag{113}$$

$$d\varepsilon_{\text{V}}^{\text{S}} = \frac{dJ_{\text{S}}}{J_{\text{S}}} \tag{114}$$

$$d\varepsilon_{\text{V}}^{\text{FR}} = -\frac{d\rho^{\text{FR}}}{\rho_0^{\text{FR}}} \tag{115}$$

where $d\varepsilon_{\text{V}}^{\text{S}} = \text{tr}\, d\boldsymbol{\varepsilon}^{\text{S}}$ is the infinitesimal increment of volumetric strain of solid phase. Based on relations (42)-(44) the infinitesimal increment of Jacobian determinants can be written as

$$\frac{dJ_{\text{S}}}{J_{\text{S}}} = -\frac{d\rho^{\text{S}}}{\rho^{\text{S}}} \tag{116}$$

$$\frac{dJ_{\text{SN}}}{J_{\text{SN}}} = -\frac{dn^{\text{S}}}{n^{\text{S}}} \tag{117}$$

$$\frac{dJ_{\text{SR}}}{J_{\text{SR}}} = -\frac{d\rho^{\text{SR}}}{\rho^{\text{SR}}} \tag{118}$$

By substituting equations (116)-(118) into (54), the relation for the infinitesimal increment of volumetric strains is obtained

$$d\varepsilon_{\text{V}}^{\text{S}} = d\varepsilon_{\text{V}}^{\text{SR}} + d\varepsilon_{\text{V}}^{\text{SN}}. \tag{119}$$

For the solid phase, the deviatoric part of the infinitesimal increment of strain can be written as

$$d\boldsymbol{\varepsilon}^{\text{SD}} = d\boldsymbol{\varepsilon}^{\text{S}} - \frac{1}{3} d\varepsilon_{\text{V}}^{\text{S}} \mathbf{I}. \tag{120}$$

In the following, for all phases a homogeneous and isotropic state and a zero initial stress in the reference configuration are assumed. It is worth remarking that the assumption implies that $\mathbf{u}_{\text{S}0} = \mathbf{0}$, $J_{\text{SN0}} = 1$, $J_{\text{SR0}} = 1$, $J_{\text{S0}} = 1$ and $J_{\text{F0}} = 1$.

The first step is to introduce the linearization of the decoupled constitutive relations (89), (93), (98) and (100). Assume the material deforms only by a small amount from the reference configuration ($n^{\text{S}} = n_0^{\text{S}}$), since $\varepsilon_{\text{V}}^{\text{SN}} \approx \frac{J_{\text{SN}}}{J_{\text{SN0}}} - 1 = \frac{n_0^{\text{S}}}{n^{\text{S}}} - 1$ is a small quantity, equation (100) can be expanded into a Taylor series, which gives

$$P + p^{\text{FR}} = -\frac{\partial \left(P + p^{\text{FR}}\right)}{\partial n^{\text{S}}}\bigg|_{n^{\text{S}} = n_0^{\text{S}}} \left(\frac{n_0^{\text{S}}}{n^{\text{S}}} - 1\right) + o(\frac{n_0^{\text{S}}}{n^{\text{S}}} - 1) = \rho^{\text{S}} \frac{\partial^2 \tilde{\psi}_3^{\text{S}}\left(n^{\text{S}}\right)}{\partial n^{\text{S}} \partial n^{\text{S}}}\bigg|_{n^{\text{S}} = n_0^{\text{S}}} \left(\frac{n_0^{\text{S}}}{n^{\text{S}}} - 1\right) + o(\frac{n_0^{\text{S}}}{n^{\text{S}}} - 1)$$

$$= \rho^{\text{S}} \frac{\partial^2 \tilde{\psi}_3^{\text{S}}\left(n^{\text{S}}\right)}{\frac{\partial J_{\text{SN}}}{J_{\text{SN}}} \frac{\partial J_{\text{SN}}}{J_{\text{SN}}}}\bigg|_{J_{\text{SN}} = J_{\text{SN0}}} \left(\frac{J_{\text{SN}}}{J_{\text{SN0}}} - 1\right) + o(\frac{J_{\text{SN}}}{J_{\text{SN0}}} - 1) = \rho^{\text{S}} \frac{\partial^2 \tilde{\psi}_3^{\text{S}}\left(n^{\text{S}}\right)}{\partial \varepsilon_{\text{V}}^{\text{SN}} \partial \varepsilon_{\text{V}}^{\text{SN}}}\bigg|_{\varepsilon_{\text{V}}^{\text{SN}} = 0} \varepsilon_{\text{V}}^{\text{SN}} + o(\varepsilon_{\text{V}}^{\text{SN}})$$

$$\tag{121}$$

where $o(\cdot)$ denotes a term that approaches zero faster than $(\cdot)$.

Neglecting the higher-order terms $o(\varepsilon_V^{SN})$ in equation (121), the linearized constitutive relation related to $\varepsilon_V^{SN}$ is obtained as

$$P + p^{FR} = K^{SN}\varepsilon_V^{SN} \tag{122}$$

where $K^{SN}$ is a material constant defined as

$$K^{SN} = \rho^S \left.\frac{\partial^2 \tilde{\psi}_3^S(n^S)}{\partial \varepsilon_V^{SN} \partial \varepsilon_V^{SN}}\right|_{\varepsilon_V^{SN}=0}. \tag{123}$$

Analogously to the linearization process discussed above, the constitutive relations (93) and (98) can also be expanded into a Taylor series as

$$p^{SR} = -\rho^{SR} \left.\frac{\partial p^{SR}}{\partial \rho^{SR}}\right|_{\rho^{SR}=\rho_0^{SR}} \left(\frac{\rho_0^{SR}}{\rho^{SR}}-1\right) + o\left(\frac{\rho_0^{SR}}{\rho^{SR}}-1\right) = -\rho^{SR} \left.\frac{\partial^2 \psi^{SR}(\rho^{SR})}{\partial \rho^{SR} \partial \rho^{SR}}\right|_{\rho^{SR}=\rho_0^{SR}} \left(\frac{\rho_0^{SR}}{\rho^{SR}}-1\right) + o\left(\frac{\rho_0^{SR}}{\rho^{SR}}-1\right)$$

$$= -\rho^{SR} \left.\frac{\partial^2 \psi^{SR}(\rho^{SR})}{\frac{\partial J_{SR}}{J_{SR}} \frac{\partial J_{SR}}{J_{SR}}}\right|_{J_{SR}=J_{SR0}} \left(\frac{J_{SR}}{J_{SR0}}-1\right) + o\left(\frac{J_{SR0}}{J_{SR}}-1\right) = -\rho^{SR} \left.\frac{\partial^2 \psi^{SR}(\rho^{SR})}{\partial \varepsilon_V^{SR} \partial \varepsilon_V^{SR}}\right|_{\varepsilon_V^{SR}=0} \varepsilon_V^{SR} + o(\varepsilon_V^{SR})$$

$$\tag{124}$$

$$p^{FR} = -\rho^{FR} \left.\frac{\partial p^{FR}}{\partial \rho^{FR}}\right|_{\rho^{FR}=\rho_0^{FR}} \left(\frac{\rho_0^{FR}}{\rho^{FR}}-1\right) + o\left(\frac{\rho_0^{FR}}{\rho^{FR}}-1\right) = -\rho^{FR} \left.\frac{\partial^2 \psi^{FR}(\rho^{FR})}{\partial \rho^{FR} \partial \rho^{FR}}\right|_{\rho^{FR}=\rho_0^{FR}} \left(\frac{\rho_0^{FR}}{\rho^{FR}}-1\right) + o\left(\frac{\rho_0^{FR}}{\rho^{FR}}-1\right)$$

$$= -\rho^{FR} \left.\frac{\partial^2 \psi^{FR}(\rho^{FR})}{\partial \varepsilon_V^{FR} \partial \varepsilon_V^{FR}}\right|_{\varepsilon_V^{FR}=0} \varepsilon_V^{FR} + o(\varepsilon_V^{FR})$$

$$\tag{125}$$

Neglecting the terms of order $o(\varepsilon_V^{SR})$ and $o(\varepsilon_V^{FR})$, the linearized version of constitutive relations (124) and (125) can be expressed as

$$p^{SR} = -K^{SR}\varepsilon_V^{SR} \tag{126}$$

$$p^{FR} = -K^{FR}\varepsilon_V^{FR} \tag{127}$$

where $K^{SR}$ and $K^{FR}$ are the material constants defined as

$$K^{SR} = \rho^{SR} \left.\frac{\partial^2 \psi^{SR}(\rho^{SR})}{\partial \varepsilon_V^{SR} \partial \varepsilon_V^{SR}}\right|_{\varepsilon_V^{SR}=0} \tag{128}$$

$$K^{FR} = \rho^{FR} \left. \frac{\partial^2 \psi^{FR}(\rho^{FR})}{\partial \varepsilon_V^{FR} \partial \varepsilon_V^{FR}} \right|_{\varepsilon_V^{FR}=0} . \tag{129}$$

It is worth remarking that equations (126) and (127) which model the behavior of intrinsic materials are both established at the microscopic scale.

Consider the constitutive relation (89) which is only related to the deviatoric part of the Cauchy stress tensor. The invariants of $\breve{\mathbf{C}}_S$, $\breve{\mathbf{b}}_S$, $\mathbf{C}_S$ and $\mathbf{b}_S$ are

$$I_{\breve{\mathbf{C}}_S} = I_{\breve{\mathbf{b}}_S} = J_S^{-2/3} I_{\mathbf{C}_S} = J_S^{-2/3} I_{\mathbf{b}_S} \tag{130}$$

$$II_{\breve{\mathbf{C}}_S} = II_{\breve{\mathbf{b}}_S} = J_S^{-4/3} II_{\mathbf{C}_S} = J_S^{-4/3} II_{\mathbf{b}_S} \tag{131}$$

$$III_{\breve{\mathbf{C}}_S} = III_{\breve{\mathbf{b}}_S} = 1, \quad III_{\mathbf{C}_S} = III_{\mathbf{b}_S} = J_S^2 . \tag{132}$$

As $III_{\breve{\mathbf{C}}_S} = 1$, the free energy related to $\breve{\mathbf{C}}_S$ can be expressed in terms of invariants as

$$\tilde{\psi}_1^S(\breve{\mathbf{C}}_S) = \tilde{\psi}_1^S(I_{\breve{\mathbf{b}}_S}, II_{\breve{\mathbf{b}}_S}). \tag{133}$$

With the introduction of the Helmholtz free energy function based on the invariants in (133), the constitutive relation (89) can be rewritten as

$$\mathbf{T}^{SD} = 2\rho^S \mathbf{F}_S \frac{\partial \tilde{\psi}_1^S(I_{\breve{\mathbf{b}}_S}, II_{\breve{\mathbf{b}}_S})}{\partial I_{\breve{\mathbf{b}}_S}} \frac{\partial I_{\breve{\mathbf{b}}_S}}{\partial \mathbf{C}_S} \mathbf{F}_S^T + 2\rho^S \mathbf{F}_S \frac{\partial \tilde{\psi}_1^S(I_{\breve{\mathbf{b}}_S}, II_{\breve{\mathbf{b}}_S})}{\partial II_{\breve{\mathbf{b}}_S}} \frac{\partial II_{\breve{\mathbf{b}}_S}}{\partial \mathbf{C}_S} \mathbf{F}_S^T \tag{134}$$

where

$$\frac{\partial I_{\breve{\mathbf{b}}_S}}{\partial \mathbf{C}_S} = J_S^{-2/3} \mathbf{I} - \frac{1}{3} I_{\mathbf{b}_S} J_S^{-2/3} \mathbf{C}^{-1} \tag{135}$$

$$\frac{\partial II_{\breve{\mathbf{b}}_S}}{\partial \mathbf{C}_S} = -\frac{2}{3} J_S^{-4/3} II_{\mathbf{b}_S} \mathbf{I} + J_S^{-4/3} I_{\mathbf{b}_S} \mathbf{b}_S - J_S^{-4/3} \mathbf{b}_S^2 . \tag{136}$$

Substitution of equations (135) and (136) into (134) leads for the first term on the right hand side of equation (134)

$$2\rho^S \mathbf{F}_S \frac{\partial \tilde{\psi}_1^S(I_{\breve{\mathbf{b}}_S}, II_{\breve{\mathbf{b}}_S})}{\partial I_{\breve{\mathbf{b}}_S}} \frac{\partial I_{\breve{\mathbf{b}}_S}}{\partial \mathbf{C}_S} \mathbf{F}_S^T = 2 J_S^{-2/3} \rho^S \frac{\partial \tilde{\psi}_1^S(I_{\breve{\mathbf{b}}_S}, II_{\breve{\mathbf{b}}_S})}{\partial I_{\breve{\mathbf{b}}_S}} \left( -\frac{1}{3} I_{\mathbf{b}_S} \mathbf{I} + \mathbf{b}_S \right) \tag{137}$$

and for the second term on the right hand of equation (134)

$$2\rho^S \mathbf{F}_S \frac{\partial \tilde{\psi}_1^S(I_{\breve{\mathbf{b}}_S}, II_{\breve{\mathbf{b}}_S})}{\partial II_{\breve{\mathbf{b}}_S}} \frac{\partial II_{\breve{\mathbf{b}}_S}}{\partial \mathbf{C}_S} \mathbf{F}_S^T = 2 J_S^{-4/3} \rho^S \frac{\partial \tilde{\psi}_1^S(I_{\breve{\mathbf{b}}_S}, II_{\breve{\mathbf{b}}_S})}{\partial II_{\breve{\mathbf{b}}_S}} \left( -\frac{2}{3} II_{\mathbf{b}_S} \mathbf{I} + I_{\mathbf{b}_S} \mathbf{b}_S - \mathbf{b}_S^2 \right). \tag{138}$$

Substitution of equations (137) and (138) into (134) leads to the following expression for the deviatoric part of the Cauchy stress

$$\mathbf{T}^{SD} = \alpha_0 \mathbf{I} + \alpha_1 \mathbf{b}_S + \alpha_2 \mathbf{b}_S^2 \tag{139}$$

where

$$\alpha_0\left(I_{\mathbf{b}_S}, II_{\mathbf{b}_S}, III_{\mathbf{b}_S}\right) = -\frac{2}{3}\rho^S J_S^{-2/3} I_{\bar{\mathbf{b}}_S} \frac{\partial \tilde{\psi}_1^S\left(I_{\bar{\mathbf{b}}_S}, II_{\bar{\mathbf{b}}_S}\right)}{\partial I_{\bar{\mathbf{b}}_S}} - \frac{4}{3}\rho^S J_S^{-4/3} II_{\bar{\mathbf{b}}_S} \frac{\partial \tilde{\psi}_1^S\left(I_{\bar{\mathbf{b}}_S}, II_{\bar{\mathbf{b}}_S}\right)}{\partial II_{\bar{\mathbf{b}}_S}} \quad (140)$$

$$\alpha_1\left(I_{\mathbf{b}_S}, II_{\mathbf{b}_S}, III_{\mathbf{b}_S}\right) = 2\rho^S J_S^{-2/3} \frac{\partial \tilde{\psi}_1^S\left(I_{\bar{\mathbf{b}}_S}, II_{\bar{\mathbf{b}}_S}\right)}{\partial I_{\bar{\mathbf{b}}_S}} + 2\rho^S J_S^{-4/3} I_{\bar{\mathbf{b}}_S} \frac{\partial \tilde{\psi}_1^S\left(I_{\bar{\mathbf{b}}_S}, II_{\bar{\mathbf{b}}_S}\right)}{\partial II_{\bar{\mathbf{b}}_S}} \quad (141)$$

$$\alpha_2\left(I_{\mathbf{b}_S}, II_{\mathbf{b}_S}, III_{\mathbf{b}_S}\right) = -2\rho^S J_S^{-4/3} \frac{\partial \tilde{\psi}_1^S\left(I_{\bar{\mathbf{b}}_S}, II_{\bar{\mathbf{b}}_S}\right)}{\partial II_{\bar{\mathbf{b}}_S}}. \quad (142)$$

The linearization of the constitutive relation with the general representation (139) was also carried out by Kelly (2020) and leads

$$\mathbf{T}^{SD} = \lambda^{SD}\left(\operatorname{tr}\boldsymbol{\varepsilon}^S\right)\mathbf{I} + 2G\boldsymbol{\varepsilon}^S \quad (143)$$

where

$$\alpha = \alpha_0 + \alpha_1 + \alpha_2 \quad (144)$$

$$\lambda^{SD} = 2\left(\frac{\partial \alpha}{\partial I_{\mathbf{b}_S}} + 2\frac{\partial \alpha}{\partial II_{\mathbf{b}_S}} + \frac{\partial \alpha}{\partial III_{\mathbf{b}_S}}\right)_{\mathbf{b}_S = \mathbf{I}} = -\frac{4}{3}\rho^S \frac{\partial \tilde{\psi}_1^S\left(I_{\bar{\mathbf{b}}_S}, II_{\bar{\mathbf{b}}_S}\right)}{\partial I_{\bar{\mathbf{b}}_S}} - \frac{4}{3}\rho^S \frac{\partial \tilde{\psi}_1^S\left(I_{\bar{\mathbf{b}}_S}, II_{\bar{\mathbf{b}}_S}\right)}{\partial II_{\bar{\mathbf{b}}_S}} \quad (145)$$

$$G = \left(\alpha_1 + 2\alpha_2\right)_{\mathbf{b}_S = \mathbf{I}} = 2\rho^S \frac{\partial \tilde{\psi}_1^S\left(I_{\bar{\mathbf{b}}_S}, II_{\bar{\mathbf{b}}_S}\right)}{\partial I_{\bar{\mathbf{b}}_S}} + 2\rho^S \frac{\partial \tilde{\psi}_1^S\left(I_{\bar{\mathbf{b}}_S}, II_{\bar{\mathbf{b}}_S}\right)}{\partial II_{\bar{\mathbf{b}}_S}}. \quad (146)$$

From (145) and (146) it can be concluded that $\lambda^{SD} = -2G/3$. Considering

$$\boldsymbol{\varepsilon}^S = \boldsymbol{\varepsilon}^{SD} + \frac{1}{3}\left(\operatorname{tr}\boldsymbol{\varepsilon}^S\right)\mathbf{I} \quad (147)$$

the constitutive equation (143) for the deviatoric part of the stress tensor can be expressed as

$$\mathbf{T}^{SD} = 2G\boldsymbol{\varepsilon}^{SD} \quad (148)$$

Equations (122), (126), (127) and (148) are the linearized versions of the decoupled constitutive equations (100), (93), (98) and (89) which includes the three volumetric moduli $K^{SN}$, $K^{SR}$ and $K^{FR}$, and the shear modulus $G$. It is found that the decoupled constitutive relations are the same as those derived by Carrol and Katsube (1983) based on the pure macroscopic theory. However, these equations are not easy to use because the volumetric part of the strain tensor $\boldsymbol{\varepsilon}^S$ cannot be expressed explicitly.

In the second step the coupled linearized constitutive equations will be established based on the above results. Firstly, it can be noted that equations (107) and (110) are respectively the same as equations (89) and (98). Consequently, equations (148) and (127) are also the linearized versions of (107) and (110), respectively. Then, by applying an analogous linearization process to the decoupled constitutive relations, the linearized version of equations (108) and (109) can be written as

$$\begin{bmatrix} P+p^{FR} \\ -n_0^S p^{FR} \end{bmatrix} = \begin{bmatrix} K_{11} & K_{12} \\ K_{21} & K_{22} \end{bmatrix} \begin{bmatrix} \varepsilon_V^S \\ \varepsilon_V^{SR} \end{bmatrix} \quad (149)$$

where

$$\left(n_0^S - n^S \varepsilon_V^{SN}\right) p^{FR} \approx n_0^S p^{FR} \quad (150)$$

is adopted because $n^S \varepsilon_V^{SN} \ll n_0^S$ in the linearized form.

The material constants in equation (149) can be expressed as

$$K_{11} = \rho^S \frac{\partial^2 \left[\tilde{\psi}_3^S\left(n^S\right) + \psi^{SR}\left(\rho^{SR}\right)\right]}{\partial \varepsilon_V^S \partial \varepsilon_V^S}\bigg|_{\varepsilon_V^S = 0} \quad (151)$$

$$K_{12} = \frac{\partial \left[\tilde{\psi}_3^S\left(n^S\right) + \psi^{SR}\left(\rho^{SR}\right)\right]}{\partial \varepsilon_V^S \partial \varepsilon_V^{SR}}\bigg|_{\varepsilon_V^{SR} = 0, \varepsilon_V^S = 0} \quad (152)$$

$$K_{21} = \frac{\partial^2 \left[\tilde{\psi}_3^S\left(n^S\right) + \psi^{SR}\left(\rho^{SR}\right)\right]}{\partial \varepsilon_V^{SR} \partial \varepsilon_V^S}\bigg|_{\varepsilon_V^S = 0, \varepsilon_V^{SR} = 0} \quad (153)$$

$$K_{22} = \frac{\partial \left[\tilde{\psi}_3^S\left(n^S\right) + \psi^{SR}\left(\rho^{SR}\right)\right]}{\partial \varepsilon_V^{SR} \partial \varepsilon_V^{SR}}\bigg|_{\varepsilon_V^{SR} = 0}. \quad (154)$$

The expression of the volumetric part of free energy for solid phase per unit volume based on the linearized constitutive equations (121) and (126) is considered

$$\rho^S \left[\tilde{\psi}_3^S\left(n^S\right) + \psi^{SR}\left(\rho^{SR}\right)\right] = \frac{1}{2} K^{SN} \left(\varepsilon_V^{SN}\right)^2 + \frac{1}{2} n^S K^{SR} \left(\varepsilon_V^{SR}\right)^2 \quad (155)$$

where $n^S$ is the volume fraction in the current configuration, which can be expressed based on Eqs. (112) and (117) as

$$n^S = n_0^S - n^S \varepsilon_V^{SN}. \quad (156)$$

By substituting of equation (156) into (155), the last term on the right hand of equation (155) is rewritten as

$$\frac{1}{2} n^S K^{SR} \left(\varepsilon_V^{SR}\right)^2 = \frac{1}{2} n_0^S K^{SR} \left(\varepsilon_V^{SR}\right)^2 - \frac{1}{2} n^S K^{SR} \left(\varepsilon_V^{SR}\right)^2 \varepsilon_V^{SN}. \quad (157)$$

It is found that the last term on the right hand side of (157) is a small quantity compared with the first term. With considering the assumption of linearity, equation (155) can be expressed as

$$\rho^S \left[\tilde{\psi}_3^S\left(n^S\right) + \psi^{SR}\left(\rho^{SR}\right)\right] \approx \frac{1}{2} K^{SN} \left(\varepsilon_V^{SN}\right)^2 + \frac{1}{2} n_0^S K^{SR} \left(\varepsilon_V^{SR}\right)^2. \quad (158)$$

By introducing (119) into (158), the free energy per unit volume at the macroscopic scale can be rewritten as

$$\rho^S\left[\tilde{\psi}_3^S\left(n^S\right)+\psi^{SR}\left(\rho^{SR}\right)\right] \approx \frac{1}{2}K^{SN}\left(\varepsilon_V^S\right)^2 - K^{SN}\varepsilon_V^S\varepsilon_V^{SR} + \frac{1}{2}\left(n_0^S K^{SR}+K^{SN}\right)\left(\varepsilon_V^{SR}\right)^2. \qquad (159)$$

Substituting (159) into (151)-(154) yields

$$K_{11} = K^{SN} \qquad (160)$$

$$K_{12} = K_{21} = -K^{SN} \qquad (161)$$

$$K_{22} = n_0^S K^{SR} + K^{SN} \qquad (162)$$

Considering the coupled behavior between $\varepsilon_V^S$ and $\varepsilon_V^{SR}$, the linearized version of coupled constitutive equations (108) and (109) can be expressed by the following matrix representation

$$\begin{bmatrix} P+p^{FR} \\ -n_0^S p^{FR} \end{bmatrix} = \begin{bmatrix} K^{SN} & -K^{SN} \\ -K^{SN} & n_0^S K^{SR}+K^{SN} \end{bmatrix} \begin{bmatrix} \varepsilon_V^S \\ \varepsilon_V^{SR} \end{bmatrix}. \qquad (163)$$

Equation (163) can also be rewritten as

$$\begin{bmatrix} \varepsilon_V^S \\ \varepsilon_V^{SR} \end{bmatrix} = \begin{bmatrix} \dfrac{n_0^S K^{SR}+K^{SN}}{n_0^S K^{SR} K^{SN}} & \dfrac{1}{n_0^S K^{SR}} \\ \dfrac{1}{n_0^S K^{SR}} & \dfrac{1}{n_0^S K^{SR}} \end{bmatrix} \begin{bmatrix} P+p^{FR} \\ -n_0^S p^{FR} \end{bmatrix} = \begin{bmatrix} \dfrac{1}{K^S} & \dfrac{1}{n_0^S K^{SR}} \\ \dfrac{1}{n_0^S K^{SR}} & \dfrac{1}{n_0^S K^{SR}} \end{bmatrix} \begin{bmatrix} P+p^{FR} \\ -n_0^S p^{FR} \end{bmatrix} \qquad (164)$$

In equation (164), the bulk modulus of the solid skeleton $K^S$ can be obtained under drained condition, i.e. the real fluid pressure $p^{FR}=0$, as

$$K^S = \frac{n_0^S K^{SR} K^{SN}}{n_0^S K^{SR}+K^{SN}} \qquad (165)$$

Equations (148), (127) and (164) form the linearized version of the complete macro-microscopic coupled constitutive relations. The macro-microscopic coupled constitutive relations can also be represented as

$$\boldsymbol{\varepsilon}^{SD} = \frac{\mathbf{T}^{SD}}{2G} \qquad (166)$$

$$\varepsilon_V^S = \frac{1}{K^S}\left(P+p^{FR}\right) - \frac{1}{K^{SR}}p^{FR} = \frac{1}{K^S}\left[P+\left(1-\frac{K^S}{K^{SR}}\right)p^{FR}\right] \qquad (167)$$

$$\varepsilon_V^{SR} = \frac{1}{n_0^S K^{SR}}\left(P+p^{FR}\right) - \frac{1}{K^{SR}}p^{FR} = -\frac{1}{K^{SR}}p^{SR} \qquad (168)$$

$$\varepsilon_V^{FR} = -\frac{p^{FR}}{K^{FR}}. \qquad (169)$$

By combining (148) and (163), the following matrix representation of the coupled constitutive relations can be obtained

$$\begin{bmatrix} t_{xx} + p^{FR} \\ t_{yy} + p^{FR} \\ t_{zz} + p^{FR} \\ t_{xy} \\ t_{xz} \\ t_{yz} \\ -n_0^S p^{FR} \end{bmatrix} = \begin{bmatrix} K^{SN} + \frac{4}{3}G & K^{SN} - \frac{2}{3}G & K^{SN} - \frac{2}{3}G & 0 & 0 & 0 & -K^{SN} \\ K^{SN} - \frac{2}{3}G & K^{SN} + \frac{4}{3}G & K^{SN} - \frac{2}{3}G & 0 & 0 & 0 & -K^{SN} \\ K^{SN} - \frac{2}{3}G & K^{SN} - \frac{2}{3}G & K^{SN} + \frac{4}{3}G & 0 & 0 & 0 & -K^{SN} \\ 0 & 0 & 0 & 2G & 0 & 0 & 0 \\ 0 & 0 & 0 & 0 & 2G & 0 & 0 \\ 0 & 0 & 0 & 0 & 0 & 2G & 0 \\ -K^{SN} & -K^{SN} & -K^{SN} & 0 & 0 & 0 & n_0^S K^{SR} + K^{SN} \end{bmatrix} \begin{bmatrix} \varepsilon_{xx} \\ \varepsilon_{yy} \\ \varepsilon_{zz} \\ \varepsilon_{xy} \\ \varepsilon_{xz} \\ \varepsilon_{yz} \\ \varepsilon_V^{SR} \end{bmatrix} \quad (170).$$

Herrin the components of the symmetric stress tensor and strain tensor in matrix form read

$$\mathbf{T} = \begin{bmatrix} t_{xx} & t_{xy} & t_{xz} \\ t_{xy} & t_{yy} & t_{yz} \\ t_{xz} & t_{yz} & t_{zz} \end{bmatrix} \tag{171}$$

$$\boldsymbol{\varepsilon}^S = \begin{bmatrix} \varepsilon_{xx} & \varepsilon_{xy} & \varepsilon_{xz} \\ \varepsilon_{xy} & \varepsilon_{yy} & \varepsilon_{yz} \\ \varepsilon_{xz} & \varepsilon_{yz} & \varepsilon_{zz} \end{bmatrix}. \tag{172}$$

Equation (170) can also be rewritten as

$$\mathbf{T} + p^{FR}\mathbf{I} = \mathbf{C}^S : \boldsymbol{\varepsilon}^S - K^{SN} \varepsilon_V^{SR} \mathbf{I} \tag{173}$$

$$-n_0^S p^{FR} = -K^{SN} \boldsymbol{\varepsilon}^S : \mathbf{I} + \left(n_0^S K^{SR} + K^{SN}\right) \varepsilon_V^{SR} \tag{174}$$

where

$$\mathbf{C}^S = 2G \overset{4}{\mathbf{I}} + \left(K^{SN} - \frac{2}{3}G\right) \mathbf{I} \otimes \mathbf{I}. \tag{175}$$

Based on the definition of $\varepsilon_V^{SN}$ in equation (112) and relation(119), the volume fraction $n^S$ of the solid material is obtained

$$n^S = \frac{n_0^S}{1 + \varepsilon_V^S - \varepsilon_V^{SR}} \tag{176}$$

# 6 Comparisons of the results obtained with other models

## 6.1 Comparison with Biot's model (Biot, 1941; Biot and Willis, 1957)

Biot's model for the porous media can be expressed as (Detournay and Cheng, 1993)

$$\mathbf{T}^{SD} = 2G \boldsymbol{\varepsilon}^{SD} \tag{177}$$

$$\varepsilon_V^S = -\frac{1}{K^S}\left[P - \alpha p^{FR}\right] \tag{178}$$

$$\zeta = -\frac{\alpha}{K^S}\left(P - \frac{p^{FR}}{B}\right). \tag{179}$$

In equations (179), factor $\zeta$ is related to the variation in fluid content (refer to Cheng, 2016) defined as

$$\frac{\partial \zeta}{\partial t} + \text{div}\left(n^F \mathbf{v}_{FS}\right) = 0 \tag{180}$$

where $\alpha$ and $B$ are the material constants.

In order to compare the present model with Biot's model, equation (5) is substituted into (25) which leads to the following equation of mass conservation of the fluid phase as

$$\rho^{FR} \frac{d^S n^F}{dt} + n^F \frac{d^S \rho^{FR}}{dt} + n^F \rho^{FR} \text{div} \mathbf{v}_S + n^F \rho^{FR} \text{div} \mathbf{v}_{FS} + \rho^{FR} \mathbf{v}_{FS} \cdot \text{grad} n^F + n^F \mathbf{v}_{FS} \cdot \text{grad} \rho^{FR} = 0 \tag{181}$$

By substituting

$$n^F \text{div} \mathbf{v}_{FS} + \mathbf{v}_{FS} \text{grad} n^F = \text{div}\left(n^F \mathbf{v}_{FS}\right) \tag{182}$$

into (181), the mass conservation of the fluid phase can be written as

$$\rho^{FR} \frac{d^S n^F}{dt} + n^F \frac{d^S \rho^{FR}}{dt} + n^F \rho^{FR} \text{div} \mathbf{v}_S + \rho^{FR} \text{div}\left(n^F \mathbf{v}_{FS}\right) + n^F \mathbf{v}_{FS} \cdot \text{grad} \rho^{FR} = 0. \tag{183}$$

The last term on the left hand side of equation (183) can be neglected, i.e., the product of a velocity and the gradient of the density is small compared with the fourth term (Verruijt, 2008). By substituting (2), (54) and (180) into (183) and after some manipulations, equation (183) can be rewritten as

$$-\frac{1}{\rho^S} \frac{d^S \rho^S}{dt} + n^F \frac{1}{\rho^{FR}} \frac{d^S \rho^{FR}}{dt} + n^S \frac{1}{\rho^{SR}} \frac{d^S \rho^{SR}}{dt} - \frac{\partial \zeta}{\partial t} = 0. \tag{184}$$

By introducing the definition of strain-like variables and cancelling the small strain items compared with 1, the linearized form of equation (184) can be rewritten as

$$\varepsilon_V^S - n_0^S \varepsilon_V^{SR} - \left(1 - n_0^S\right) \varepsilon_V^{FR} - \zeta = 0. \tag{185}$$

Substituting the constitutive equations (127), (167) and (168) into (185) yields

$$\zeta = \frac{1}{K^S}\left[P + \left(1 - \frac{K^S}{K^{SR}}\right) p^{FR}\right] - n_0^S \left[\frac{1}{n_0^S K^{SR}}\left(P + p^{FR}\right) - \frac{1}{K^{SR}} p^{FR}\right] + \left(1 - n_0^S\right) \frac{p^{FR}}{K^{FR}}$$

$$= \left(\frac{1}{K^S} - \frac{1}{K^{SR}}\right) P + \left(\frac{1}{K^S} + \frac{1 - n_0^S}{K^{FR}} - \frac{2 - n_0^S}{K^{SR}}\right) p^{FR} \tag{186}$$

By comparing equations (148), (167) and (186) with (177), (178) and (179), it is found that all the three equations in Biot's model can be derived based on the proposed model. The difference lies in that in continuum mechanics the traction force is taken to be positive and compression force is taken to be negative, whereas the hydrostatic pressure is positive under compression. The relations among the material constants in the proposed model and Biot's model are

$$\alpha = \frac{K^S}{K^{SR}} \tag{187}$$

$$B = \frac{K^{FR}\left(K^{SR} - K^S\right)}{K^{FR} K^{SR} + \left(1 - n_0^S\right) K^S K^{SR} - \left(2 - n_0^S\right) K^S K^{FR}}. \tag{188}$$

It can be concluded that the macroscopic constitutive relations of Biot's model can be obtained based on the linearized version of the proposed model, and thus the gap between the theory of porous media and Biot's theory is bridged. However, it is worth noting that the intrinsic constitutive relations of the constituents are not considered in Biot's model. Although both models exploit three volumetric moduli, the present model gives the evolution of the porosity by considering the volumetric strain of the solid skeleton and the volumetric strain of the real solid material.

## 6.2 Comparison with the Lopatnikov and Cheng model (Lopatnikov and Cheng, 2002)

A constitutive model for compressible fluid-saturated porous media in which the evolution of porosity is taken into account has been proposed by Lopatnikov and Cheng (2002). The comparison between the present model with the model by Lopatnikov and Cheng shows that $\mathbf{T}^{SD}$ can be interpreted as the deviatoric part of total external stress tensor, $p^{SR}$ is the solid mean compressive stress, $p^{FR}$ is the fluid pressure, $\boldsymbol{\varepsilon}^{SD}$ is the average external solid deviatoric strain, $\varepsilon_V^{SR}$ is the average internal solid volumetric strain, and $\varepsilon_V^{FR}$ is the average fluid volumetric strain. The Lopatnikov and Cheng model for an isothermal process can be expressed as

$$\mathbf{T}^{SD} = 2G\boldsymbol{\varepsilon}^{SD} \tag{189}$$

$$p^{SR} = -K^{SR}\varepsilon_V^{SR} - K^{Interaction}\Delta n^F \tag{190}$$

$$p^{FR} = -K^{FR}\varepsilon_V^{FR} \tag{191}$$

$$p^{SR} - p^{FR} = -\left(1-n_0^F\right)K^{Porosity}\Delta n^F - \left(1-n_0^F\right)K^{Interaction}\varepsilon_V^{SR}. \tag{192}$$

In the model by Lopatnikov and Cheng, a term "micro-inhomogeneity" is introduced to denote the coupled behavior of Helmholtz free energy $\tilde{\psi}_3^S(n^S)$ and $\tilde{\psi}_4^S(\rho^{SR})$, and the volumetric modulus $K^{Interaction}$ is associated with the micro-inhomogeneity. If the Helmholtz free energy $\tilde{\psi}_2^S(n^S, \rho^{SR})$ can be decoupled into $\tilde{\psi}_3^S(n^S)$ and $\tilde{\psi}_4^S(\rho^{SR})$, the volumetric modulus $K^{Interaction}$ which is related to the coupled behavior between $\varepsilon_V^{SR}$ and $(n^F - n_0^F)$ should vanish. In this case, equations (189), (190) and (191) are generally the same as the linearized version of the decoupled constitutive relations (148), (126) and (127), respectively. For the comparison with equation (179), the following relations based on equations (112) and (70) are introduced

$$n^F - n_0^F = n_0^S \varepsilon_V^{SN} \tag{193}$$

$$-p^{SR} + p^{FR} = \frac{1}{n_0^S}\left(P + p^{FR}\right) \tag{194}$$

where $n_0^S \approx n^S$ is taken into account in equations (193) and (194). Substituting (193) and (194) into

(122) yields

$$-p^{\mathrm{SR}} + p^{\mathrm{FR}} = K^{\mathrm{SN}} \frac{n^{\mathrm{F}} - n_0^{\mathrm{F}}}{\left(n_0^{\mathrm{S}}\right)^2} \tag{195}$$

It can be readily proved that equation (195) is the same as (192) by introducing

$$K^{\mathrm{SN}} = \left(n_0^{\mathrm{S}}\right)^3 K^{\mathrm{porosity}} \tag{196}$$

It should be noted that the Lopatnikov and Cheng model is also obtained based on the decoupled form of the present linearized model for the micro-homogeneous saturated porous media. Compared with the Lopatnikov and Cheng model, the macro-microscopic coupled form of the proposed model is more suitable for engineering practice.

## 6.3 Comparison with the Serpieri and Rosati model (Serpieri and Rosati, 2011; Serpieri, 2011)

Serpieri and Rosati (2011) also proposed a general form of constitutive relations for the fluid-saturated porous media. The linearized expression of the Serpieri and Rosati model is represented by Serpieri (2011). In this section, only the linearized versions of the proposed model and the Serpieri and Rosati model are compared. The linearized expression of Serpieri and Rosati model (Serpieri, 2011) can be written as

$$\mathbf{T}^{\varepsilon^{\mathrm{SD}}} = 2G\boldsymbol{\varepsilon}^{\mathrm{SD}} \tag{197}$$

$$\begin{bmatrix} -p^{\varepsilon_{\mathrm{V}}^{\mathrm{S}}} \\ -n_0^{\mathrm{S}} p^{\mathrm{FR}} \end{bmatrix} = \begin{bmatrix} \bar{K}_0 & \bar{K}_0^{\bar{\varepsilon}_v, \hat{\varepsilon}_v} \\ \bar{K}_0^{\bar{\varepsilon}_v, \hat{\varepsilon}_v} & \bar{K}_0^{\hat{\varepsilon}_v, \hat{\varepsilon}_v} \end{bmatrix} \begin{bmatrix} \varepsilon_{\mathrm{V}}^{\mathrm{S}} \\ \varepsilon_{\mathrm{V}}^{\mathrm{SR}} \end{bmatrix} \tag{198}$$

$$p^{\mathrm{FR}} = K^{\mathrm{FR}} \varepsilon_{\mathrm{V}}^{\mathrm{FR}} \tag{199}$$

where the quantities $\mathbf{T}^{\varepsilon^{\mathrm{SD}}}$ and $-p^{\varepsilon_{\mathrm{V}}^{\mathrm{S}}}$ are interpreted as the deviatoric and volumetric part of the effective stress (Travascio et al., 2015; Serpieri and Travascio, 2017). Taking $\mathbf{T}^{\varepsilon^{\mathrm{SD}}}$ as $\mathbf{T}^{\mathrm{SD}}$ and $-p^{\varepsilon_{\mathrm{V}}^{\mathrm{S}}}$ as $P + p^{\mathrm{FR}}$, it is found that the constitutive relations (197), (198) and (199) are the same as the equations (148), (149) and (127). If the Helmholtz free energy $\tilde{\psi}_2^{\mathrm{S}}\left(n^{\mathrm{S}}, \rho^{\mathrm{SR}}\right)$ can be decoupled into $\tilde{\psi}_3^{\mathrm{S}}\left(n^{\mathrm{S}}\right)$ and $\tilde{\psi}_4^{\mathrm{S}}\left(\rho^{\mathrm{SR}}\right)$, the material constants in the Serpieri and Rosati model satisfy

$$\bar{K}_0 = K^{\mathrm{SN}} \tag{200}$$

$$\bar{K}_0^{\bar{\varepsilon}_v, \hat{\varepsilon}_v} = -K^{\mathrm{SN}} \tag{201}$$

$$\bar{K}_0^{\hat{\varepsilon}_v, \hat{\varepsilon}_v} = n_0^{\mathrm{S}} K^{\mathrm{SR}} + K^{\mathrm{SN}}. \tag{202}$$

The main difference lies in the way that, in the presented model, only three volumetric moduli are required, whereas, in the Serpieri and Rosati model, four volumetric moduli are used. This is due to the fact that the decoupled form of free energy is not considered in the work by Serpieri and Rosati (2011). Based on the previous discussions, it can be concluded that the macro-microscopic coupled model (148), (163) and (127) is suitable for the micro-homogeneous material, the Serpieri and Rosati model (197)-(199) is suitable for the micro-inhomogeneous material.

# 7. On effective stress of saturated porous media

## 7.1 Effective stress related to the deformation of saturated porous media

The effective stress related to the deformation can be defined as the stress that governs the stress-strain behavior of saturated porous media. For the sake of simplicity, the following discussion is restricted to the assumption of small strains and only mechanical power is considered. The dissipation inequality (79) can be rewritten in the local form as

$$\left(\mathbf{T} + p^{FR}\mathbf{I}\right) : \frac{d\mathbf{\varepsilon}^S}{dt} - n_0^S p^{FR} \frac{d\varepsilon_V^{SR}}{dt} - n_0^F p^{FR} \frac{d\varepsilon_V^{FR}}{dt} - \rho^S \frac{d\psi^S}{dt} - \rho^F \frac{d\psi^F}{dt} \geq 0. \quad (203)$$

Based on the Coleman-Noll approach and the linearization method, the constitutive relations obtained by equation (203) represent the linearized version of the macro-microscopic coupled constitutive relations.

The effective stress has been defined as the work conjugate to the strain of solid skeleton (Houlsby, 1979; Borja, 2006). It has been pointed out that any linear combination of the stress variables is acceptable as the work conjugate to the strain of solid skeleton (Houlsby, 1997). For example, the dissipation inequality (203) can be rewritten as

$$\mathbf{T} : \frac{d\mathbf{\varepsilon}^S}{dt} + p^{FR}\left(\frac{d\varepsilon_V^S}{dt} - n_0^S \frac{d\varepsilon_V^{SR}}{dt} - n_0^F \frac{d\varepsilon_V^{FR}}{dt}\right) - \rho^S \frac{d\psi^S}{dt} - \rho^F \frac{d\psi^F}{dt} \geq 0. \quad (204)$$

Substituting (185) into (204) yields

$$\mathbf{T} : \frac{d\mathbf{\varepsilon}^S}{dt} + p^{FR} \frac{d\zeta}{dt} - \rho^S \frac{d\psi^S}{dt} - \rho^F \frac{d\psi^F}{dt} \geq 0 \quad (205)$$

Using the Coleman-Noll approach and the linearization method, the constitutive relations obtained from equation (205) are those of Biot.

It can be found that the work conjugate to the strain of solid skeleton is $\mathbf{T} + p^{FR}\mathbf{I}$ and $\mathbf{T}$ in (203) and (205), respectively. From the proposed macro-microscopic coupled constitutive relations and the model by Biot, it can be found that both $\mathbf{T} + p^{FR}\mathbf{I}$ and $\mathbf{T}$ can be adopted as stress state variables. For a complete description of the stress-strain behavior of saturated porous media, however, the constitutive relations should include all the terms required. In other words, the coupled behavior between $\mathbf{\varepsilon}^S$ and $\varepsilon_V^{SR}$ in (203) or the coupled behavior between $\mathbf{\varepsilon}^S$ and $\zeta$ in (205) should be taken into account.

When the focus is only on the deformation of solid skeleton, the effective stress can be defined as a combination of the externally applied stresses and the pressure of fluid phase. This enables the transformation of the stress-strain behavior of a porous media into a mechanically equivalent single-phase continuum (Nuth and Laloui, 2010). From equation (167) the effective stress can be expressed as

$$\mathbf{T}^{e}_{\text{deformation}} = \mathbf{T} + \left(1 - \frac{K^{S}}{K^{SR}}\right) p^{FR}\mathbf{I} = 2G\boldsymbol{\varepsilon}^{SD} + K^{S}\varepsilon_{V}^{S}. \tag{206}$$

It can therefore be concluded that Biot's effective stress is suitable for describing the deformation of the solid skeleton. A complete description of the mechanical behavior of porous media, however, should consider the coupled behavior of all state variables involved.

### 7.2 Effective stress related to strength of saturated porous media

Although the mechanical behavior of the proposed model is purely elastic the corresponding limit , i.e. the strength, is of practical importance. In this context, only the yield stress and yield criterion are discussed in this section. Enhanced concepts in plasticity such as, the hardening modes and the flow rules, are not taken into account in the present paper. For the isotropic porous media it assumes that failure occurs at particular combinations of the variables related to the current stress state, which permits the formulation of a yield criterion. Considering the macro-microscopic coupled constitutive relations, the function of the yield criterion should include the stress variables $\mathbf{T} + p^{FR}\mathbf{I}$ and $p^{FR}$, i.e.

$$f = f\left(\mathbf{T} + p^{FR}\mathbf{I}, p^{FR}\right) \tag{207}$$

If it is assumed that the real solid material behaves plastic incompressible, the saturated porous media exhibits volumetric yield only when $P + p^{FR}$ reaches a critical level. Thus, the failure criterion can be expressed as

$$f = f\left(\mathbf{T} + p^{FR}\mathbf{I}\right) \tag{208}$$

Equation (208) indicates that the onset of yielding of saturated porous media can also be described by using the effective stress formulation given by Terzaghi.

The idea that the strength is governed by Terzaghi's effective stress and the deformation of solid skeleton is governed by Biot's effective stress has been widely discussed by several authors, e.g. Skempton (1961); Nur and Byerlee (1971); Jaeger et al. (2007). An extension of the proposed model to the plastic range with the introduction of an enhanced failure criteria, such as the Mohr-Coulomb criterion or the unified strength theory (Yu, 2004), needs further research studies.

## 8. Comparisons with experimental results

### 8.1 Comparison with stress-strain data of Weber sandstone

The linearized version of the proposed model is expressed by stresses in equations (166)-(169). In particular, equation (166) denotes the deviatoric part of the constitutive relation, equations (168) and (169) denote the volumetric part of the constitutive relations of real solid material and real fluid material, respectively. These constitutive relations discussed in Section 6.1 are widely accepted in the literature. In order to verify the volumetric part of the proposed constitutive model equation (167) this is compared with the experimental data of conventional compression tests carried out in a triaxial apparatus for the dry and fluid saturated Weber sandstone (Nur and Byerlee, 1971). The data measured are the volume strain $\varepsilon_V^S$ depending on the applied total mean pressure $P$ and the pore fluid pressure $p^{FR}$. To approximate the experimental data, the following constitutive relations for an elastic material behavior are considered:

$$\varepsilon_V^S = \frac{1}{K^S} P \tag{209}$$

$$\varepsilon_V^S = \frac{1}{K^S}\left(P + p^{FR}\right) \tag{210}$$

$$\varepsilon_V^S = \frac{1}{K^S}\left(P + p^{FR}\right) - \frac{1}{K^{SR}} p^{FR} = \frac{1}{K^S}\left[P + \left(1 - \frac{K^S}{K^{SR}}\right)p^{FR}\right]. \tag{211}$$

From Fig. 2 it is clearly visible that for the same applied pressure $P$ the measured volume strain is different for the dry and the fluid saturated material. With respect to equation (209) the linear approximation of the data for the dry specimen, leads the bulk modulus $K^S$ of the solid skeleton. For higher applied pressures the experimental data deviate from the approximated straight line which indicates that for Weber sandstone the value of $K^S$ is pressure dependent. Figure 3 shows the volume strain depending on the difference between the confining pressure $P$ and the pore fluid pressure $p^{FR}$ according to the effective stress law proposed by Terzaghi (1936). The experimental data are compared with the constitutive equation (210). Figure 4 shows that the scatter of the experimental data can be improved to a certain extent when the compressibility of the real solid material, i.e. the bulk modulus $K^{SR}$ of the real solid material, is taken into account in accordance with equation (211). Equation (211) assumes that the volumetric strain of the solid skeleton is governed by Biot's effective stress, which is the same as the one for the linearized version of the proposed model. It can therefore be concluded that Biot's effective stress is more suitable to be defined as the effective stress related to stress-strain behavior for fluid saturated porous media. The comparison also indicates that three volumetric moduli adopted in the linearized version of the proposed model are sufficient to describe the substantial mechanical behavior of Weber sandstone.

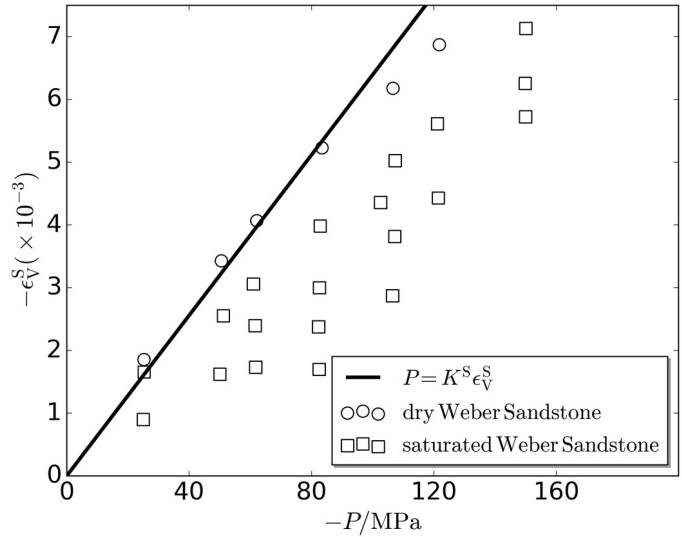

Fig. 2 Volumetric strain of Weber sandstone versus total mean pressure: shapes denote experimental

data from Nur and Byerlee (1971); the solid line is obtained from the constitutive relation (209).

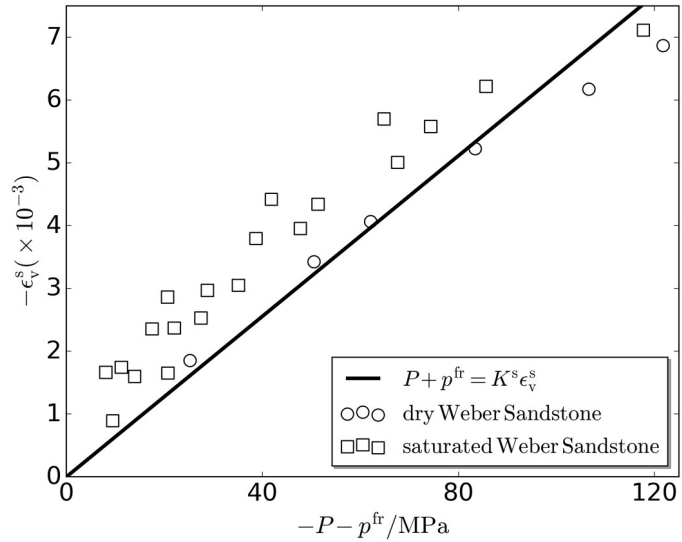

Fig. 3 Volumetric strain of Weber sandstone versus confining Terzaghi's effective stress: shapes denote experimental data from Nur and Byerlee (1971); the solid line is obtained from the constitutive relation (210).

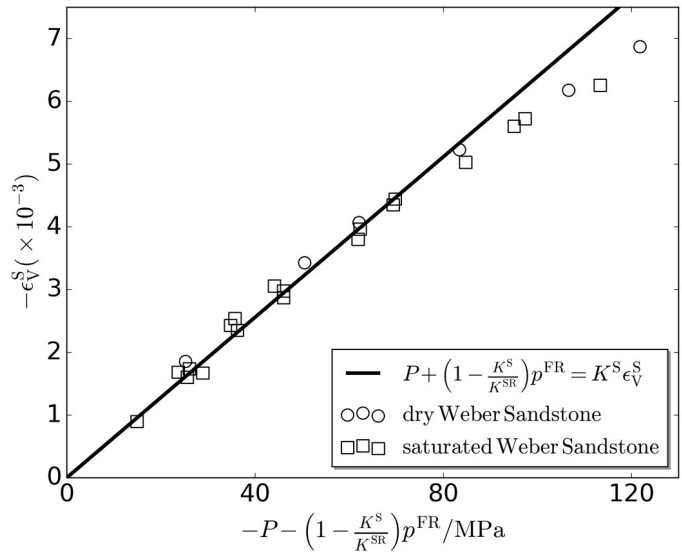

Fig. 4 Volumetric strain of Weber sandstone versus-Biot's effective stress: shapes denote experimental data from Nur and Byerlee (1971); the solid line is obtained from the constitutive relation (211).

## 8.2 Comparison with strength data of Darley Daly sandstone

In this section the strength data of Darley Daly sandstone tested by Murrel (1965) under conventional triaxial stress condition are considered. For five different confining stresses $t_1$, the axial stress $t_3$ versus the fluid pressure $p^{FR}$ is shown in Fig. 5. It is obvious that the failure stress $t_1$ strongly depends on the amount of the confining stress. The experimental data can be approximated by five distinct lines.

The alternative representation, i.e. $t_1 + p^{FR}$ versus $t_3 + p^{FR}$, shown in Fig. 6 indicates, that the experimental data for all confining stresses nearly form a single failure curve, which in a first approximation, can be well related to the effective stress by Terzaghi as also discussed in Section 7.2.

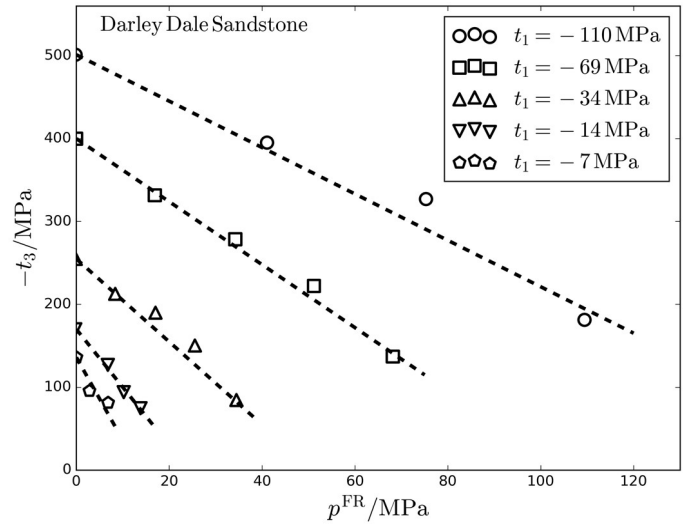

Fig.5 Strength data of Darley Dale sandstone on the $p^{FR}$ versus $t_3$ plane (experimental data from Murrell, 1965; redrawn from Jaeger et al., 2007).

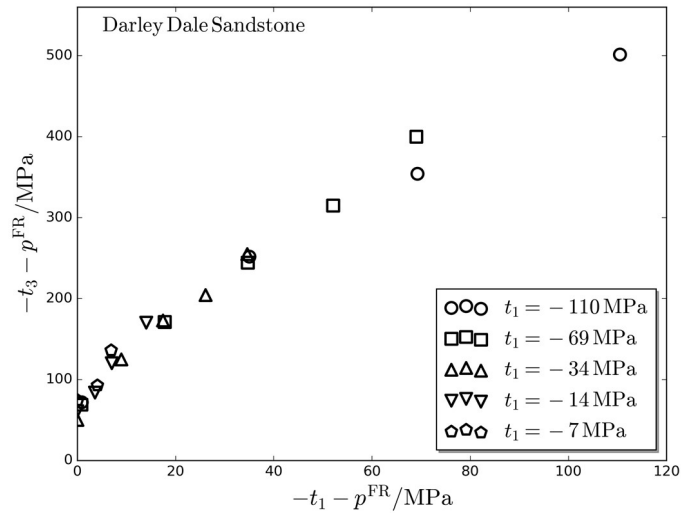

Fig.6 Strength data of Darley Dale sandstone on the $t_1 + p^{FR}$ versus $t_3 + p^{FR}$ plane (experimental data from Murrell, 1965; redrawn from Jaeger et al., 2007).

# 9. Conclusions

In this paper, the macro-microscopic coupled constitutive relations for the fluid-saturated porous media with compressible constituents have been derived within the framework of porous media theory. Two different sets of independent variables are introduced to implement the coupled behavior between the compressibility of the solid skeleton and the real solid material. It is found that the mechanical behavior of saturated porous media with compressible constituents may be governed by the deviatoric part of the right Cauchy-Green deformation tensor, the partial density of solid phase, the density of the real solid material, the density of the real fluid material, as well as the relative velocity of fluid phase.

The linearized version of the constitutive model is discussed and compared with three similar models published by other authors. In particular, it is found that Biot's model can also be derived based on the linearized version of the constitutive model developed in the present paper. This result allows the conclusion that with the proposed model the gap between the theory of porous media and Biot's theory is bridged. A common feature shared by the present linearized model and Biot's model is that both models exploit three volumetric moduli. Compared with Biot's model, the present model takes the evolution of the porosity into account based on the volumetric strain of the solid skeleton and the volumetric strain of the real solid material.

The comparison between the linearized constitutive model and the Lopatnikov and Cheng model indicates that the Lopatnikov and Cheng model is the same as the linearized version of the proposed model for micro-homogeneous saturated porous media, however, the proposed model is more general for a refined practical application. Compared with the linearized formulations derived by Serpieri and Rosati, the main difference lies in the number of volumetric moduli of the solid phase. It can be concluded that the proposed model is suitable for micro-homogeneous saturated porous media and the Serpieri and Rosati model is suitable for micro-inhomogeneous saturated porous media.

## References

Anand, L.: A large deformation poroplasticity theory for microporous polymeric materials. Journal of the Mechanics and Physics of Solids. 98, 126-155 (2017)

Berryman, J.G.: Comparison of upscaling methods in poroelasticity and its generalizations. Journal of Engineering Mechanics. 131(9), 928-936 (2005). https://doi.org/10.1061/(ASCE)0733-9399(2005)131:9(928)

Biot, M.A.: General theory of three-dimensional consolidation. Journal of Applied Physics. 12, 155–164 (1941)

Biot, M.A.: Theory of deformation of a porous viscoelastic anisotropic solid. Journal of Applied Physics, 27(5), 459-467 (1956)

Biot, M.A., Willis, D.G.: The elastic coefficients of the theory of consolidation. Journal of Applied Mechanics. 15, 594-601 (1957)

Biot, M.A.: Theory of Finite Deformations of Pourous Solids. Indiana Univ. Math. J. 21(7), 597-620 (1972)

Biot, M.A., Variational Lagrangian-thermodynamics of nonisothermal finite strain mechanics of porous solids and thermomolecular diffusion. International Journal of Solids and Structures. 13, 579–597 (1977)

Bluhm, J., de Boer, R.: The volume fraction concept in the porous media theory. ZAMM-Journal of Applied Mathematics and Mechanics. 77(8), 563-577 (1997). https://doi.org/10.1002/zamm.19970770803.

de Boer, R.: Highlights in the historical development of the porous media theory: toward a consistent macroscopic theory. Appl. Mech. Rev. 49, 201–262 (1996). https://doi.org/10.1115/1.3101926.

de Boer, R.: Theory of porous media: highlights in historical development and current state. Springer, Berlin (2000)

de Boer, R.: Trends in continuum mechanics of porous media. Springer, Berlin (2005)

de Boer, R., Bluhm, J.: Phase transitions in gas-and liquid-saturated porous solids. Transport in porous media, 34(1-3), 249-267 (1999). https://doi.org/10.1023/A:1006577828659.

de Boer, R., Ehlers, W.: Uplift, friction and capillarity: three fundamental effects for liquid-saturated porous solids. International Journal of Solids and Structures. 26(1), 43-57 (1990). https://doi.org/10.1016/0020-7683(90)90093-B.

Borja, R.I.: Conservation laws for three-phase partially saturated granular media. In: Unsaturated Soils: Numerical and Theoretical Approaches. Springer, Berlin, Heidelberg (2005)

Borja, R.I.: On the mechanical energy and effective stress in saturated and unsaturated porous continua. International Journal of Solids and Structures. 43(6), 1764-1786 (2006). https://doi.org/10.1016/j.ijsolstr.2005.04.045.

Borja, R.I., Koliji, A.: On the effective stress in unsaturated porous continua with double porosity. Journal of the Mechanics and Physics of Solids. 57(8),1182-1193 (2009). https://doi.org/10.1016/j.jmps.2009.04.014.

Bowen, R.M.: Incompressible porous media models by use of the theory of mixtures. International Journal of Engineering Science. 18(9), 1129-1148 (1980) https://doi.org/10.1016/0020-7225(80)90114-7

Bowen, R.M.: Compressible porous media models by use of the theory of mixtures. International Journal of Engineering Science. 20(6), 697-735 (1982). https://doi.org/10.1016/0020-7225(82)90082-9.

Bowen, R. M.: Introduction to Continuum Mechanics for Engineers. Plenum Press, New York (1989)

Carroll, M.M., Katsube, N.: The role of Terzaghi effective stress in linearly elastic deformation. J. Energy Res. Tech. 105, 509-511 (1983). https://doi.org/10.1115/1.3230964

Chen, X., Hicks, M. A.: A constitutive model based on modified mixture theory for unsaturated rocks. Computers and Geotechnics. 38(8), 925-933 (2011). https://doi.org/10.1016/j.compgeo.2011.04.008

Cheng, A.H.D.: Poroelasticity. Springer International Publishing, Cham (2016).

Coleman, B.D., Noll, W.: The thermodynamics of elastic materials with heat conduction and viscosity. Archive for Rational Mechanics & Analysis. 13(1), 167-178 (1963).

Coussy, O., Dormieux, L., & Detournay, E. (1998). From mixture theory to Biot's approach for porous media. International Journal of Solids and Structures, 35(34-35), 4619-4635.

Coussy, O.: Poromechanics. John Wiley & Sons, Chichester (2004).

Detournay, E., Cheng, A.H.D.: Fundamentals of poroelasticity. In Analysis and design methods (pp. 113-171). Pergamon (1993).

Drass, M., Schneider, J., Kolling, S.: Novel volumetric Helmholtz free energy function accounting for isotropic cavitation at finite strains. Materials & Design. 138, 71-89 (2018). https://doi.org/10.1016/j.matdes.2017.10.059.

Ehlers, W.: Foundations of multiphasic and porous materials. In Porous media (pp. 3-86). Springer, Berlin, Heidelberg 2002.

Ehlers, W.: Challenges of porous media models in geo-and biomechanical engineering including electro-chemically active polymers and gels. International Journal of Advances in Engineering Sciences and Applied Mathematics. 1(1), 1-24 (2009). https://doi.org/10.1007/s12572-009-0001-z.

Ehlers, W.: Effective Stresses in Multiphasic Porous Media: A thermodynamic investigation of a fully non-linear model with compressible and incompressible constituents. Geomechanics for Energy and the Environment, 15, pp.35-46. (2018) https://doi.org/10.1016/j.gete.2017.11.004.

Fillunger P (1936), Erdbaumechanik?, Selbstverlag des Verfassers, Wien.

Flory, P.J.: Thermodynamic relations for high elastic materials. Transactions of the Faraday Society. 57,

829-838. (1961)

Gajo, A.: A general approach to isothermal hyperelastic modelling of saturated porous media at finite strains with compressible solid constituents. Proceedings of the Royal Society A: Mathematical, Physical and Engineering Sciences, 466(2122), 3061-3087(2010)

Holzapfel, A.G.: Nonlinear solid mechanics II. John Wiley, Chichester (2000).

Hornung, U.: Homogenization and Porous Media. Springer, New York (1997)

Houlsby, G.T.: The work input to a granular material. Géotechnique. 29(3), 354-358 (1979) https://doi.org/10.1680/geot.1979.29.3.354.

Houlsby, G.T.: The work input to an unsaturated granular material. Géotechnique. 47(1), 193-196 (1997). https://doi.org/10.1680/geot.1997.47.1.193.

Hu Y.Y.: Study on the super viscoelastic constitutive theory for saturated porous media. Applied Mathematics and Mechanics, 37(6), 584-598 (2016). (in Chinese)

Hu Y.Y.: Thermodynamics-based constitutive theory for unsaturated porous rock. Journal of Zhejiang University (Engineering Edition), 51( 2), 255- 263 (2017). (in Chinese)

Hu Y.Y.: Isothermal hyperelastic model for saturated porous media based on poromechanics In: Proceedings of China-Europe Conference on Geotechnical Engineering, SSGG, edited by W. Wu and H.-S. Yu, 31~34 (2018).

Jaeger, J. C., Cook, N. G., Zimmerman, R.: Fundamentals of rock mechanics (4$^{th}$ edition). Blackwell Publishing, Malden (2007).

Kelly, P.A.: Mechanics Lecture Notes: An introduction to Solid Mechanics. (2020), Available from http://homepages.engineering.auckland.ac.nz/~pkel015/SolidMechanicsBooks/index.html

Lade, P.V., de Boer, R.D.: The concept of effective stress for soil, concrete and rock. Geotechnique, 47(1), 61-78 (1997)

Laloui, L., Klubertanz, G., Vulliet, L.: Solid–liquid–air coupling in multiphase porous media. International Journal for Numerical and Analytical Methods in Geomechanics. 27(3), 183-206 (2003). https://doi.org/10.1002/nag.269

Liu, I.S.: A solid–fluid mixture theory of porous media. International Journal of Engineering Science. 84, 133-146 (2014). https://doi.org/10.1016/j.ijengsci.2014.07.002.

Lewis, R., and Schrefler, B.: The Finite Element Method in the Static and Dynamic Deformation and Consolidation of Porous Media. John and Wiley (1998).

Lopatnikov, S.L., Cheng, A.D.: Variational formulation of fluid infiltrated porous material in thermal and mechanical equilibrium. Mechanics of Materials. 34(11), 685-704 (2002). https://doi.org/10.1016/S0167-6636(02)00168-0.

Lopatnikov, S.L., Cheng, A.D.: Macroscopic Lagrangian formulation of poroelasticity with porosity dynamics. Journal of the Mechanics and Physics of Solids. 52(12), 2801-2839 (2004). https://doi.org/10.1016/j.jmps.2004.05.005

Lopatnikov, S.L., Gillespie, J.W.: Poroelasticity-I: governing equations of the mechanics of fluid-saturated porous materials. Transport in porous media. 84(2), 471-492 (2010). https://doi.org/10.1007/s11242-009-9515-x

Mosler, J., Bruhns, O.T.: Towards variational constitutive updates for non-associative plasticity models at finite strain: models based on a volumetric-deviatoric split. International Journal of Solids and Structures. 46(7-8), 1676-1684 (2009). https://doi.org/10.1016/j.ijsolstr.2008.12.008.

Müller I.: Thermodynamics, Pitman, Boston (1985)

Murrell, S.A.F: The effect of triaxial stress systems on the strength of rocks at atmospheric temperatures.

Geophysical Journal International, 10(3), 231-281 (1965)

Nur, A., Byerlee, J.D.: An exact effective stress law for elastic deformation of rock with fluids. Journal of Geophysical Research. 76, 6414–6419 (1971).

Nuth, M., Laloui, L.: Effective stress concept in unsaturated soils: clarification and validation of a unified framework. International Journal for Numerical and Analytical Methods in Geomechanics. 32(7), 771-801 (2010)

Passman, S.L., Nunziato, J.W., Walsh, E.K.: A theory of multiphase mixtures. In Rational thermodynamics (pp. 286-325). Springer, New York (1984)

Rajagopal, K. R., Tao, L.: On the propagation of waves through porous solids. International Journal of Non-Linear Mechanics, 40(2-3), 373-380 (2005)

Schanz, M., Diebels, S.: A comparative study of Biot's theory and the linear Theory of Porous Media for wave propagation problems. Acta Mechanica. 161, 213–235 (2003). https://doi.org/10.1007/s00707-002-0999-5

Schanz, M.: Poroelastodynamics: linear models, analytical solutions, and numerical methods. Applied Mechanics Reviews. 62(3), 1-15 (2009). https://doi.org/10.1115/1.3090831

Serpieri, R.: A rational procedure for the experimental evaluation of the elastic coefficients in a linearized formulation of biphasic media with compressible constituents. Transport in porous media. 90(2), 479-508 (2011). https://doi.org/10.1007/s11242-011-9796-8.

Serpieri, R., Rosati, L.: Formulation of a finite deformation model for the dynamic response of open cell biphasic media. Journal of the Mechanics and Physics of Solids. 59(4), 841-862 (2011). https://doi.org/10.1016/j.jmps.2010.12.016.

Serpieri, R., Travascio, F.: Variational Continuum Multiphase Poroelasticity. Springer International Publishing AG, Singapore. (2017)

Simo, J.C.: A framework for finite strain elastoplasticity based on maximum plastic dissipation and the multiplicative decomposition: Part I. Continuum formulation. Computer Methods in Applied Mechanics and Engineering. 66(2), 199-219 (1988). https://doi.org/10.1016/0045-7825(88)90076-X.

Skempton, A.W.: Effective stress in soils, concrete and rocks. In: Proc. Conf. Pore Pressure and Suction in Soils (1961)

Terzaghi, K.: The shearing resistance of saturated soils and the angle between the planes of shear. In: 1st International Conference on Soil Mechanics and Foundation Engineering. 1, 54–56 (1936)

Travascio F., Asfour S., Serpieri R. and Rosati L., (2015). Analysis of the consolidation problem of compressible porous media by a macroscopic variational continuum approach. Mathematics and Mechanics of Solids, 22(5). https://doi.org/10.1177/1081286515616049

Verruijt, A.: Encyclopedia of Hydrological Sciences, chap. Consolidation of Soils. John Wiley, Chichester (2008). https://doi.org/10.1002/0470848944.hsa303.

Wei, C., Muraleetharan, K. K.: A continuum theory of porous media saturated by multiple immiscible fluids: II. Lagrangian description and variational structure. International Journal of Engineering Science, 40(16), 1835-1854 (2002)

Wilmanski, K.: A thermodynamic model of compressible porous materials with the balance equation of porosity. Transport in Porous Media, 32(1), 21-47 (1998). https://doi.org/10.1023/A:1006563932061.

Wilmanski, K.: Continuum Thermodynamics-Part I: Foundations (Vol. 1). World Scientific (2008)

Yu.: Unified Strength theory and its applications. Springer, Berlin (2004)

Zhang, Y.: Mechanics of adsorption–deformation coupling in porous media. Journal of the Mechanics and Physics of Solids, 114, 31-54 (2018)